\documentclass[12pt,a4paper,onecolumn,draftcls]{IEEEtran}
\usepackage{epsfig,epstopdf,graphics,subfigure,psfrag,amsmath,cases}
\usepackage{latexsym,amssymb,amsmath,epsfig,subfigure,algorithm,amsthm}
\usepackage{algorithmic}
\usepackage{color}
\usepackage{url}
\usepackage{scrtime}
\usepackage{bm}
\usepackage{graphicx}
\usepackage{hyperref}
\usepackage{subfigure}
\usepackage{bbding}
\usepackage{multicol}
\usepackage{graphics}
\usepackage{amsthm}
\usepackage{cite}
\usepackage{array}
\usepackage[OT1]{fontenc}
\usepackage{booktabs}

\author{Yuanxin Cai,~\IEEEmembership{Graduate Student Member,~IEEE,} Zhiqiang Wei,~\IEEEmembership{Member,~IEEE,}\\
Shaokang Hu,~\IEEEmembership{Student Member,~IEEE,} Chang Liu,~\IEEEmembership{Member,~IEEE,}\\
Derrick Wing Kwan Ng,~\IEEEmembership{Fellow,~IEEE,} and Jinhong Yuan,~\IEEEmembership{Fellow,~IEEE}
\thanks{
Y. Cai is with the Key Laboratory of Information and Communication Systems, Ministry of Information Industry, Beijing Information Science and Technology University, China and the School of Electrical Engineering and Telecommunications, the University of New South Wales, Australia
 (e-mail: carina.cyx@gmail.com). 

S. Hu, C. Liu, D. W. K. Ng, and J. Yuan are with the School of Electrical Engineering and Telecommunications, the University of New South Wales, Australia
(e-mail: \{shaokang.hu, chang.liu, w.k.ng, j.yuan\}@unsw.edu.au).

Z. Wei is with the Institute for Digital Communications (IDC),
Friedrich-Alexander University Erlangen-Nuremberg, Germany (e-mail:
zhiqiang.wei@fau.de).
This paper has been presented in part at IEEE ICC Workshops 2020 \cite{9145224}.}
}

\title{Resource Allocation and 3D Trajectory Design for Power-Efficient IRS-Assisted UAV-NOMA Communications}

{\newtheorem{Thm}{Theorem}}

\begin{document}
\maketitle

\vspace{-12mm}
\begin{abstract}
	In this paper, an intelligent reflecting surface (IRS) is introduced to assist an unmanned aerial vehicle (UAV) communication system based on non-orthogonal multiple access (NOMA) for serving multiple ground users.
	We aim to minimize the average total system energy consumption by jointly designing the resource allocation strategy, the three dimensional (3D) trajectory of the UAV, as well as the phase control at the IRS.
	The design is formulated as a non-convex optimization problem taking into account the maximum tolerable outage probability constraint and the individual minimum data rate requirement.
	To circumvent the intractability of the design problem due to the altitude-dependent Rician fading in UAV-to-user links, we adopt the deep neural network (DNN) approach to accurately approximate the corresponding effective channel gains, which facilitates the development of a low-complexity suboptimal iterative algorithm via dividing the formulated problem into two subproblems and address them alternatingly.
	Numerical results demonstrate that the proposed algorithm can converge to an effective solution within a small number of iterations and illustrate some interesting insights:
(1) IRS enables a highly flexible UAV's 3D trajectory design via recycling the dissipated radio signal for improving the achievable system data rate and reducing the flight power consumption of the UAV;
(2) IRS provides a rich array gain through passive beamforming in the reflection link, which can substantially reduce the required communication power for guaranteeing the required quality-of-service (QoS);
(3) Optimizing the altitude of UAV's trajectory can effectively exploit the outage-guaranteed effective channel gain to save the total required communication power enabling power-efficient UAV communications;
(4) NOMA communications offer higher degrees of freedom (DoF) than that of the conventional orthogonal multiple access (OMA) scheme to minimize the average power consumption via optimizing the UAV's trajectory.
\end{abstract}

\vspace{-2mm}
\section{Introduction}
\vspace{-1mm}

In recent years, the dramatic growth in the number of wireless devices and the associated demanding quality-of-services (QoS) have fueled the development of new technologies for the fifth-generation (5G) and beyond 5G (B5G) wireless networks.
Although several potential technologies, e.g. millimeter wave and massive multiple-input multiple-output (MIMO), offer some promising solutions to guarantee ubiquitous and ultra-high data rate services, e.g. \cite{8529214,9389801}, the system performance is still limited by some bottlenecks, such as overloaded traffic demand or shadowed communication links.
Fortunately, unmanned aerial vehicles (UAVs)-enabled wireless communication systems provide a feasible solution \cite{opportunities_and_challenges,9043712}, which overcome the physical restrictions of traditional terrestrial wireless systems.
Particularly, by exploiting the high mobility of the UAV, the communication performance can be improved via cruising the UAV close to the desired users.
Also, the line-of-sight (LoS) probability between the UAV and the desired ground users increases with the operating altitude of the UAV which facilitates the establishment of a strong communication link.
Thus, UAV-enabled wireless communications which UAVs serve as aerial relays \cite{9454372}, aerial base stations \cite{DBLP:journals/corr/abs-2110-04733}, etc, have drawn significant attention from both academia and industry.

Although numerous advantages of adopting UAVs have been revealed in the literature, e.g. \cite{opportunities_and_challenges,DWK_2018_solar_power_comm}, the onboard battery capacity with limited energy storage capacity of UAVs still restricts the performance of UAV-enabled communications.
To fully unleash the performance of UAV communication systems, various studies have been conducted in the literature to improve the power efficiency.
For instance, in \cite{7510870}, the authors studied the optimal deployment of multiple UAVs to minimize the total system transmit power satisfying the individual user data rate requirement simultaneously.
However, the flight power consumption of the UAV was ignored in this work which contributes a major proportion of the total system power consumption.
Besides, the authors in \cite{rotary_wing_power} minimized the total power consumption of both communication and flying via jointly optimizing UAV's trajectory and user scheduling for a rotary-wing UAV.
Yet, pure LoS wireless channels between the UAV and ground users were assumed \cite{7510870,rotary_wing_power}, which are generally invalid in practice, particularly in urban environments.
Also, a probabilistic LoS channel model for UAV-enabled data harvesting system was proposed in \cite{you2020hybrid}, which is suitable to a system with a relatively low flying altitude UAV when the shadowing effect dominates the system performance.
{On the other hand, for a relatively high altitude UAV with a clear LoS, in \cite{3d_trajectory}, the UAV's 3D trajectory was optimized taking into account a practical model with an angle-dependent Rician fading channel to maximize the minimum average collected data rate.}
However, the considered UAV communication systems in \cite{you2020hybrid,3d_trajectory} are based on time division multiple access (TDMA) scheme and their results are not applicable to a more general system supporting multiple users simultaneously.
Besides, although a non-orthogonal multiple access (NOMA)-assisted UAV communication system with Rician fading channel model was studied in \cite{9094731}, a significant portion of the system power is still dissipated in signal transmission if the channel condition of the users is unsatisfactory.
As a result, a new technology to improve the channel quality is desired for power-efficient communication provisioning.

Most recently, intelligent reflecting surface (IRS) has attracted substantial attention in the field of wireless communications as it can reshape the signal propagation environment so as to improve the system performance.
For example, beamforming and discrete phase control of IRS-assisted systems were jointly optimized to minimize the total transmit power in \cite{8683145}.
Besides, the authors in \cite{8811733} proposed a jointly optimized active beamforming at the transmitter and passive beamforming at the IRS to maximize the received signal power at desired users.
Furthermore, it is expected that deploying an IRS in UAV-enabled communication systems can help to improve the achievable data rate for ground users with a weak channel condition.
In particular, the passive beamforming controlled by the IRS can reflect the dissipated signals transmitted from the UAV to the ground users.
This unique feature not only increases the received signal strength at the desired users, but also improves the flexibility in the UAV's trajectory design.
Thus, the integration of an IRS into UAV-based communication systems has been advocated lately.
For instance, the authors in \cite{Li2019ReconfigurableIS} maximized the average achievable data rate in IRS-assisted UAV communication systems by jointly optimizing the UAV's trajectory and the phase shift control of the IRS.
Yet, this study only focused on the case of a single-user and the proposed result is not applicable to practical multi-user systems.
Also, the joint design of two-dimensional (2D) trajectory and passive beamforming was studied in \cite{9145224} for multi-user IRS-aided UAV communications assuming the availability of perfectly known channel state information (CSI), which is overly optimistic.
Moreover, the 2D trajectory designs ignored the possibility of exploiting the altitude dimension for improving the system performance.
{Besides, the 3D trajectory optimization for UAV communications was designed in \cite{DWK_2018_solar_power_comm} and \cite{9013544}, which only consider the pure LoS channel model and a constant path loss exponent, respectively.}
Furthermore, the considered system models in \cite{Li2019ReconfigurableIS} neglected the existence of a direct link between the UAV and ground users which leads to inevitable performance degradation.
In fact, a joint resource allocation, 3D trajectory design, and phase shift control for power-efficient IRS-assisted multi-user UAV communication systems are important and challenging, which has not been reported in the literature yet.

In this paper, we address the aforementioned problems.
We study the joint design of the resource allocation, UAV's 3D trajectory, and its flight velocity, as well as the phase shift control of the IRS in a practical altitude-dependent Rician fading channel for power-efficient IRS-assisted UAV-NOMA communications.
The joint design is formulated as a non-convex optimization problem to minimize the average total power consumption of the system taking into account the minimum data rate requirement of each user and the maximum tolerable outage probability constraint.
Since the formulated problem is non-convex and highly intractable, we first propose a closed-form phase control policy for the IRS.
Then, to handle the intractability caused by the altitude-dependent Rician fading channel, we employ a deep neural network (DNN) technique to approximate the outage-guaranteed effective channel gain.
Furthermore, the obtained results are exploited to serve as a building block for the design of an iterative optimization algorithm for addressing the design problem.
In particular, we divide the problem at hand into two subproblems and solve them iteratively based on the alternating optimization method.
In each iteration, a suboptimal solution of these two subproblems are obtained by the successive convex approximation (SCA) with a fast convergence.

The remainder of this paper is organized as follows.
In Section II, we present the system and channel models.
Section III provides the joint design problem formulation.
The proposed solutions are presented in Section IV.
Section V illustrates numerical results to evaluate the performance of the proposed scheme.
Finally, we conclude the paper in Section VI.

Notation: $\mathbb{R}^{M \times N}$ and $\mathbb{C}^{M \times N}$ represent the space of a $M \times N$ matrix with real and complex entries, respectively.
$\arcsin(\cdot)$ denotes the inverse trigonometric functions of $\sin(\cdot)$.
Operator $|\cdot|$ and $\|\cdot\|$ represent the absolute value and the vector norm, respectively.
$\mathbf{X}^{\mathrm T}$ and $\mathbf{X}^{\mathrm H}$ denote the transpose and the conjugate transpose of matrix $\mathbf{X}$, respectively.
$\mathbf{X} \otimes \mathbf{Y}$ denotes the Kronecker product of two matrices $\mathbf{X}$ and $\mathbf{Y}$.
$\mathcal{CN}(0,\sigma^2)$ denote a circularly symmetric complex Gaussian (CSCG) distribution with zero mean and noise variance $\sigma^2$, and $\sim$ means ``distributed as''.
$\text{diag}(x_1,\ldots,x_K)$ represents a diagonal matrix with the diagonal elements given by $\{x_1,\ldots,x_K\}$ and $[x]^+ = \max \{0,x\}$.
$\mathcal{O}(\cdot)$ represents the big-O notation.

\vspace{-2mm}
\section{System Model}
\vspace{-1mm}

\begin{figure}[t]
  \vspace*{-8mm}
  \centering
  \includegraphics[width=3.55 in]{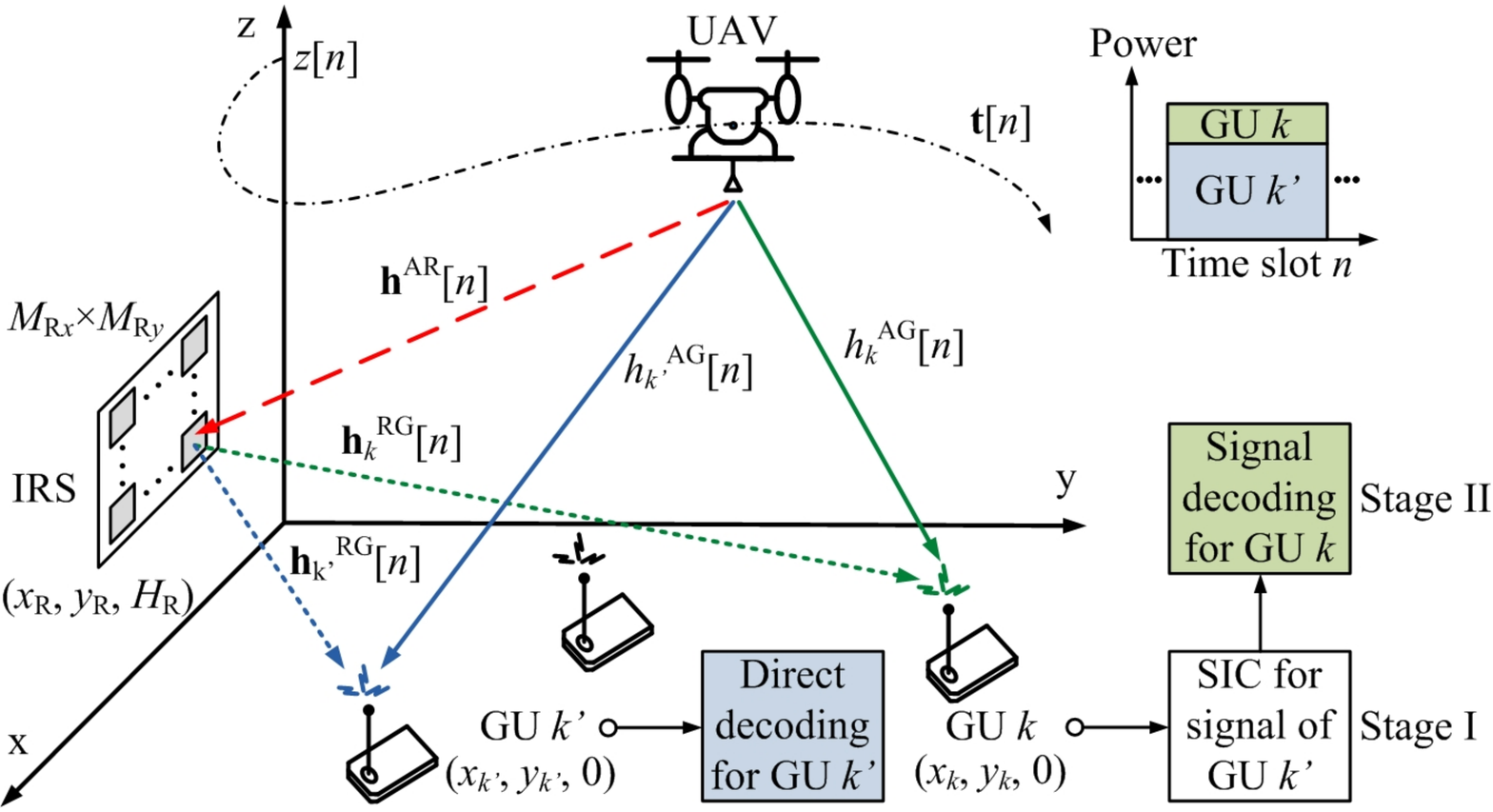}
  \vspace*{-15mm}
  \caption{An IRS-assisted UAV-NOMA communication system with multiple ground users.}
  \label{system_model}
  \vspace*{-10mm}
\end{figure}

We consider a rotary UAV-enabled downlink NOMA wireless communication system serving $K$ ground users (GUs) with the assistance of an IRS as shown in Fig. \ref{system_model}.
Particularly, the IRS is coated on the surface of a building located at the edge of the service area such that all the GUs have the opportunity to be assisted by the IRS \cite{hu2021beamforming,9557830}.
We assume that the UAV is equipped with a single antenna\footnote{Note that single-antenna UAV is commonly assumed in the literature, e.g. \cite{rotary_wing_power,lower_bound_data_rate}, to reduce the signal processing burden at the UAV.}.
Besides, the IRS consists of $M_{\mathrm Rx} \times M_{\mathrm Ry} = M_{\mathrm R} > 1$ passive reflecting elements and all the GUs are single-antenna devices.
Also, the total service time duration $T$ is divided into $N$ equal-length time slots with duration time $\tau$ (s), i.e., $T = N \tau$.
In each time slot, the UAV selects two GUs\footnote{In this paper, we consider to select two GUs to form a NOMA group since it enjoys a lower computational complexity and a shorter signal processing delay for successive interference cancellation (SIC) decoding at GUs, compared with that of grouping more NOMA users \cite{sun2018robust}.
Moreover, as shown in \cite{6868214}, the performance gain of NOMA over OMA diminishes rapidly with increasing the number of users in one NOMA group.
Therefore, the considered two-user NOMA scheme can achieve a considerable performance improvement than the conventional OMA scheme.} and serves them through NOMA.
Moreover, the UAV operates in three-dimensional (3D) space with a variable flight velocity, while the locations of all the GUs and the IRS are fixed during the whole service time, e.g. \cite{8660516,3d_trajectory}.
Also, we assume that the IRS is deployed at a high altitude above all obstacles.
The distances between the UAV and the IRS, the UAV and GU $k \in \{1, \ldots, K\}$, as well as the IRS and GU $k$ at time slot $n \in \{1,\ldots,N\}$ are given by\vspace{-2.5mm}
\begin{equation}
d^{\mathrm{AR}}[n] = \|\mathbf{l}_{\mathrm R} - \mathbf{t}[n]\|, \,\,
d_k^{\mathrm{AG}}[n] = \|\mathbf{l}_k - \mathbf{t}[n]\|, \,\, \text{and} \,\,
d_{k}^{\mathrm{RG}} = \|\mathbf{l}_{\mathrm R} - \mathbf{l}_k\|,\\[-2.5mm]
\end{equation}
respectively.
Note that $\mathbf{l}_{\mathrm R} = [x_{\mathrm R},y_{\mathrm R},H_{\mathrm R}]^{\mathrm T} \in \mathbb{R}^{3 \times 1}$, $\mathbf{l}_k = [x_k,y_k,z_k]^{\mathrm T} \in \mathbb{R}^{3 \times 1}$, and $\mathbf{t}[n] = [x[n],y[n],z[n]]^{\mathrm T} \in \mathbb{R}^{3 \times 1}$ denote the Cartesian coordinate of the IRS\footnote{Since the typical size of each element in a small-scale IRS is the same order of the wavelength of the carrier frequency, $\lambda_{\mathrm c}$, \cite{ntontin2019reconfigurable}, e.g. $\frac{\lambda_{\mathrm c}}{2}$, the separations between reflecting elements of the IRS in the $x$-dimension and the $y$-dimension, denoted as $\Delta_{\mathrm Rx}$ and $\Delta_{\mathrm Ry}$, respectively, are much shorter than that of the distance between the UAV and the IRS, $d^{\mathrm{AR}}[n]$, as well as the distance between the IRS and GUs, $d_k^{\mathrm{RG}}$.
Thus, we assume that the distance of each element of the IRS to a GU/UAV is identical, as commonly adopted in the literature \cite{Li2019ReconfigurableIS}.}, GU $k$, and the UAV at time slot $n$, respectively.

\vspace{-6mm}
\subsection{Channel Model}
\vspace{-3mm}

\begin{figure}[t]
  \vspace*{-8mm}
  \centering
  \includegraphics[width=3.4 in]{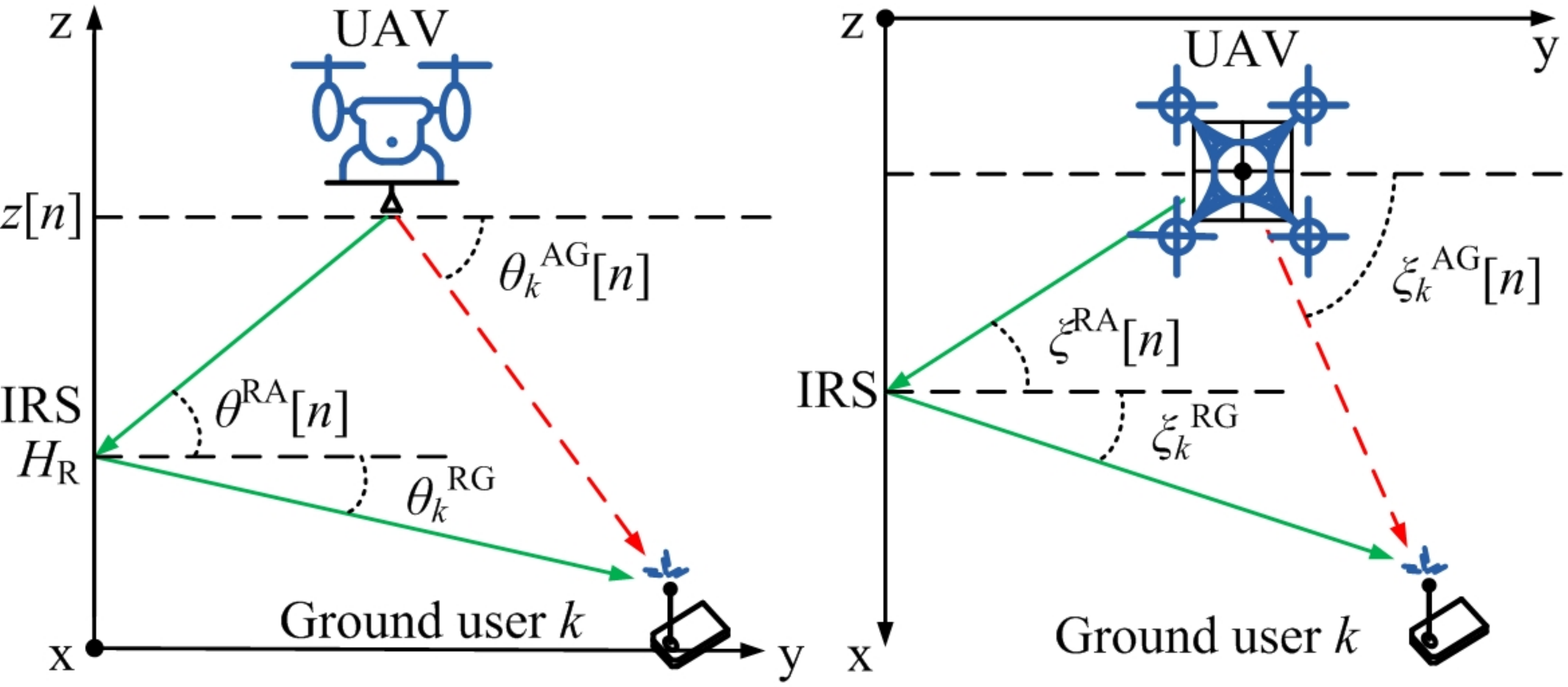}
  \vspace*{-17mm}
  \caption{The vertical and horizontal AoDs/AoAs between the UAV, IRS, and GU $k$ in the considered downlink communication system are shown on the left-hand side and the right-hand side, respectively.}
  \label{AoD}
  \vspace*{-10mm}
\end{figure}

In the considered system, we assume that the channels between the UAV and the GUs as well as the IRS and the GUs follow a frequency flat Rician fading channel model with an altitude-dependent Rician factor \cite{3d_trajectory,8255737}.
Note that the Doppler effect caused by the movement of the UAV can be well compensated by adopting existing frequency synchronization algorithms, e.g. \cite{classen1994frequency}.
{According to \cite{3d_trajectory}, the Rician factors of the direct links between the UAV and different GUs are non-identical caused by the UAV's mobility and their surrounding environments.
In fact, the altitude-dependent Rician factor for the UAV-GUs link can be modeled by an exponential function \cite{3d_trajectory,8255737}, which is given by\vspace{-2.5mm}
\begin{equation} \label{kappa_k_AG}
\kappa_k^{\mathrm{AG}}[n] = A_1 \exp \left( A_2 \theta_k^{\mathrm{AG}}[n] \right),\\[-2.5mm]
\end{equation}
where $\theta_k^{\mathrm{AG}}[n]$ is the elevation angle-of-departure (AoD) from the UAV to GU $k$ at time slot $n$, as shown in Fig. \ref{AoD}, and is given by\vspace{-2.5mm}
\begin{equation}
\theta_k^{\mathrm{AG}}[n] = \arcsin \left( \frac{z[n]}{d_k^{\mathrm{AG}}[n]} \right).\\[-2mm]
\end{equation}}
Note that $A_1>0$ and $A_2>0$ are constant parameters related to the terrain environment and can be obtained via long-term measurements.
Then, we can observe that the Rician factor is bounded by $\kappa_{\min} \leq \kappa_k^{\mathrm{AG}}[n] \leq \kappa_{\max}$, where $\kappa_{\min} = A_1$ and $\kappa_{\max} = A_1 e^{A_2 \pi/2}$.

Hence, the Rician channel between the UAV and GU $k$ at time slot $n$ is given by\vspace{-2mm}
\begin{equation} \label{h_k_AG}
h_{k}^{\mathrm{AG}}[n] = \sqrt{\frac{\beta_0}{(d_k^{\mathrm{AG}}[n])^{\alpha^{\mathrm{AG}}}}} \left[ \sqrt{\frac{\kappa_k^{\mathrm{AG}}[n]}{1 + \kappa_k^{\mathrm{AG}}[n]}} h_{k}^{\mathrm{AG,LoS}}[n] + \sqrt{\frac{1}{1 + \kappa_k^{\mathrm{AG}}[n]}} \Delta h_{k}^{\mathrm{AG}}[n] \right] \in \mathbb{C},\\[-2mm]
\end{equation}
where $h_k^{\mathrm{AG,LoS}}[n] = e^{-j 2 \pi d_k^{\mathrm{AG}[n]} / \lambda_{\mathrm c}}$ and the associated phase rotation is caused by the delay of the line-of-sight (LoS) component of UAV-GUs link, which is determined solely by their locations and is known to the system.
$\Delta h_{k}^{\mathrm{AG}}[n] \in \mathbb{C} \sim \mathcal{CN}(0,1)$ denotes the randomly scattered component of the channel experienced by GU $k$ at time slot $n$.
Note that $\beta_0\in \mathbb{R}$ and $\alpha^{\mathrm{AG}}>0$ denote the average channel power gain at the reference distance and the path loss exponent of the UAV-GUs channel, respectively.
Besides, we use $\xi_k^{\mathrm{AG}}[n]$ to represent the horizontal AoD from the UAV to GU $k$ at time slot $n$ and $\lambda_{\mathrm c}$ denotes the wavelength of the carrier frequency.
Fig. \ref{AoD} shows the geographic relations of $\sin \xi_k^{\mathrm{AG}}[n] = \frac{|x_k - x[n]|}{\sqrt{(x_k-x[n])^2 + (y_k-y[n])^2}}$ and $\cos \xi_k^{\mathrm{AG}}[n] = \frac{|y_k - y[n]|}{\sqrt{(x_k-x[n])^2 + (y_k-y[n])^2}}$.
On the other hand, the pure LoS channel\footnote{In practice, the IRS is mounted at the wall of a building that has a similar height with traditional base stations deployed in outdoor wireless communication systems \cite{ma2020enhancing}, e.g. $20-30$ meters.
Based on the field measurements in \cite{8337920,colpaert2018aerial}, the LoS probability of the air-to-air communication channel closely approaches one.
Thus, in our proposed system, we assume that the UAV-IRS link experiences the pure LoS channel which the corresponding channel coefficients can be determined by their locations.} from the UAV to the IRS at time slot $n$ is denoted as\vspace{-0.5mm}
\begin{align}
\mathbf{h}^{\mathrm{AR}}[n] &= \sqrt{\frac{\beta_0}{(d^{\mathrm{AR}}[n])^{\alpha^\mathrm{AR}}}} e^{-j \frac{2 \pi d^{\mathrm{AR}}[n]}{\lambda_{\mathrm c}}} \notag \\[-1.5mm]
&\times \big[ 1, e^{-j \frac{2 \pi \Delta_{\mathrm Rx}}{\lambda_{\mathrm c}} \sin \theta^{\mathrm{RA}}[n] \cos \xi^{\mathrm{RA}}[n]}, \ldots, e^{-j \frac{2 \pi \Delta_{\mathrm Rx}}{\lambda_{\mathrm c}} (M_{\mathrm Rx}-1) \sin \theta^{\mathrm{RA}}[n] \cos \xi^{\mathrm{RA}}[n]} \big]^{\mathrm H} \notag \\[-1.5mm]
&\otimes \big[ 1, e^{-j \frac{2 \pi \Delta_{\mathrm Ry}}{\lambda_{\mathrm c}} \sin \theta^{\mathrm{RA}}[n] \sin \xi^{\mathrm{RA}}[n]}, \ldots, e^{-j \frac{2 \pi \Delta_{\mathrm Ry}}{\lambda_{\mathrm c}} (M_{\mathrm Ry}-1) \sin \theta^{\mathrm{RA}}[n] \sin \xi^{\mathrm{RA}}[n]} \big]^{\mathrm H} \!\in\! \mathbb{C}^{M_{\mathrm R} \times 1},\\[-12mm]\notag
\end{align}
where $\alpha^{\mathrm{AR}}>0$ is the path loss exponent of the UAV-IRS link.
As shown in Fig. \ref{AoD}, $\theta^{\mathrm{RA}}[n]$ and $\xi^{\mathrm{RA}}[n]$ denote the vertical and horizontal angle-of-arrivals (AoAs) between the UAV and the IRS, respectively.
Note that $ \sin \theta^{\mathrm{RA}}[n] = \frac{|z[n] - H_{\mathrm R}|}{d^{\mathrm{AR}}[n]}$, $ \sin \xi^{\mathrm{RA}}[n] = \frac{|x_{\mathrm R} - x[n]|}{\sqrt{(x_{\mathrm R}-x[n])^2+(y_{\mathrm R}-y[n])^2}}$, and $ \cos \xi^{\mathrm{RA}}[n] = \frac{|y_{\mathrm R} - y[n]|}{\sqrt{(x_{\mathrm R}-x[n])^2+(y_{\mathrm R}-y[n])^2}}$.
Besides, the Rician channel from the IRS to GU $k$ at time slot $n$ can be modeled as\vspace{-0.5mm}
\begin{equation}
\mathbf{h}_{k}^{\mathrm{RG}}[n] = \sqrt{\frac{\beta_0}{(d_k^{\mathrm{RG}})^{\alpha^{\mathrm{RG}}}}} \left[ \sqrt{\frac{\kappa^{\mathrm{RG}}}{1+\kappa^{\mathrm{RG}}}} \mathbf{h}_{k}^{\mathrm{RG,LoS}} + \sqrt{\frac{1}{1+\kappa^{\mathrm{RG}}}} \Delta \mathbf{h}_{k}^{\mathrm{RG}} [n] \right]\in \mathbb{C}^{M_{\mathrm R} \times 1},\\[-1.5mm]
\end{equation}
where\vspace{-2mm}
\begin{align} \label{h_k_RG_LoS}
\mathbf{h}_{k}^{\mathrm{RG,LoS}} &= e^{-j \frac{2 \pi d_k^{\mathrm{RG}}}{\lambda_{\mathrm c}}} \big[ 1, e^{-j \frac{2 \pi \Delta_{\mathrm Rx}}{\lambda_{\mathrm c}} \sin \theta_{k}^{\mathrm{RG}} \cos \xi_{k}^{\mathrm{RG}}}, \ldots, e^{-j \frac{2 \pi \Delta_{\mathrm Rx}}{\lambda_{\mathrm c}} (M_{\mathrm Rx}-1) \sin \theta_{k}^{\mathrm{RG}} \cos \xi_{k}^{\mathrm{RG}}} \big]^{\mathrm H} \notag \\[-1.3mm]
&\otimes \big[ 1, e^{-j \frac{2 \pi \Delta_{\mathrm Ry}}{\lambda_{\mathrm c}} \sin \theta_{k}^{\mathrm{RG}} \sin \xi_{k}^{\mathrm{RG}}}, \ldots, e^{-j \frac{2 \pi \Delta_{\mathrm Ry}}{\lambda_{\mathrm c}} (M_{\mathrm Ry}-1) \sin \theta_{k}^{\mathrm{RG}} \sin \xi_{k}^{\mathrm{RG}}} \big]^{\mathrm H}\\[-12.5mm]\notag
\end{align}
is the LoS component that is known to the system.
$\Delta \mathbf{h}_{k}^{\mathrm{RG}}[n] \in \mathbb{C}^{M_{\mathrm R} \times 1} \sim \mathcal{CN}(0,\mathbf{I}_{M_{\mathrm R}})$, $\alpha^{\mathrm{RG}}\geq0$, and $\kappa^{\mathrm{RG}}\geq0$ represent the randomly scattered component, the path loss exponent, and the fixed Rician factor of the IRS-GUs channel, respectively.
$\theta_{k}^{\mathrm{RG}}$ and $\xi_{k}^{\mathrm{RG}}$ denote the vertical and horizontal AoDs from the IRS to GU $k$, respectively.
As shown in Fig. \ref{AoD}, we have $\sin \theta_k^{\mathrm{RG}} = \frac{H_{\mathrm R}}{d^{\mathrm{RG}}_k}$, $\sin \xi_{k}^{\mathrm{RG}} = \frac{|x_{\mathrm R} - x_k|}{\sqrt{(x_{\mathrm R}-x_k)^2+(y_{\mathrm R}-y_k)^2}}$, and $\cos \xi_{k}^{\mathrm{RG}} = \frac{|y_{\mathrm R} - y_k|}{\sqrt{(x_{\mathrm R}-x_k)^2+(y_{\mathrm R}-y_k)^2}}$.

Moreover, the IRS can manipulate the reflected signals to GUs by introducing controllable phase shifts.
The phase control matrix imposed by the IRS at time slot $n$ is given by\footnote{Note that although continuous phase control is considered in this paper, it can be extended to the case of discrete phase control via a similar approach as in \cite{hu2021robust} and with the IRS channel estimation as in \cite{liu2021deep}.}\vspace{-1mm}
\begin{equation}
\bm \Phi[n] = \text{diag} (e^{j \phi_{1,1}[n]}, \ldots,  e^{ j \phi_{m_{\mathrm Rx},m_{\mathrm Ry}}[n]}, \ldots, e^{ j \phi_{M_{\mathrm Rx},M_{\mathrm Ry}}[n]} ) \in \mathbb{C}^{M_{\mathrm R} \times M_{\mathrm R}},\\[-1mm]
\end{equation}
where $\phi_{m_{\mathrm Rx},m_{\mathrm Ry}}[n] \in [0,2 \pi)$, $m_{\mathrm Rx} = \{1, \ldots, M_{\mathrm Rx}\}$, $m_{\mathrm Ry} = \{1, \ldots, M_{\mathrm Ry}\}$, represents the phase control introduced by the $(m_{\mathrm Rx},m_{\mathrm Ry})$-th reflecting element of the IRS at time slot $n$.
Now, we define the end-to-end effective channel\footnote{The signal propagation delay between the direct link and the reflection link is negligible as it is about 2 $\mu s$ in an $500\times500$ $m^2$ service area, which is much shorter than the symbol duration in long-term
evolution systems (around 70 $\mu s$) \cite{ghosh2010fundamentals}.} between the UAV and GU $k$ at time slot $n$ as\vspace{-3mm}
\begin{equation} \label{channel}
h_{k}[n] = h_{k}^{\mathrm{AG}}[n] + (\mathbf{h}_{k}^{\mathrm{RG}}[n])^{\mathrm H} \bm \Phi[n] \mathbf{h}^{\mathrm{AR}}[n]  \in \mathbb{C}.\\[-2mm]
\end{equation}

Meanwhile, at the UAV side, the deterministic components of all channels, including the LoS components, path loss as well as the Rician factors can be determined that are available for the designed trajectory of the UAV.
Apart from the deterministic components, the distributions of the randomly scattered components in the UAV-GUs and IRS-GUs links are also available \cite{shen2019uav}.

\vspace{-3mm}
\subsection{NOMA Transmission and Achievable Data Rate}
\vspace{-1mm}

We consider NOMA transmission at the UAV to serve two GUs at each time slot as it is potential to achieve a higher power efficiency than that of the conventional OMA schemes\footnote{The proposed optimization framework is a generalized one which subsumes TDMA as a special case \cite{you2020hybrid,3d_trajectory}.} \cite{Zhiqiang_DC}.
Without loss of generality, when the UAV selects GU $k$ and GU $k'$ to form a NOMA group and instructs GU $k$ to perform SIC decoding at time slot $n$, we denote $s_{k,k'}[n]=1$, $\forall k, k'$. Otherwise, $s_{k,k'}[n]=0$.
When $s_{k,k'}[n]=1$, the UAV transmits the superimposed signals for GU $k$ and GU $k'$ simultaneously.
As illustrated in Fig. \ref{system_model}, GU $k$ is assumed as the user performing SIC which first decodes the information of GU $k'$ before decoding its own information.
Besides, GU $k'$ is assumed as the non-SIC user which directly decodes its own information while treating the interference of GU $k$ as noise.
For $s_{k,k'}[n]=1$, the achievable data rates of the two stages of SIC decoding at GU $k$ serving as a \textit{SIC user} and that of GU $k'$ serving as a \textit{non-SIC user} can be formulated as\vspace{-1mm}
\begin{align}
C_{k,k'}^{\mathrm{I,SIC}}[n] & =  \log_2 \left( 1+ \frac{p_{k'}[n] |h_{k}[n]|^2}{p_k[n] |h_{k}[n]|^2 + \sigma_{k}^2} \right), \forall n, k \neq k',\\
C_{k,k'}^{\mathrm{II,SIC}}[n] &= \log_2 \left( 1+ \frac{p_k[n] |h_{k}[n]|^2}{\sigma_k^2} \right), \forall n,k, \text{and}\label{OMA}\\
C_{k,k'}^{\mathrm{NSIC}}[n] & = \log_2 \left( 1+ \frac{p_{k'}[n] |h_{k'}[n]|^2}{p_k[n] |h_{k'}[n]|^2 + \sigma_{k'}^2} \right), \forall n, k \neq k', \\[-11mm]\notag
\end{align}
respectively, where $p_{k}[n]$, $p_{k'}[n]$, $\sigma_{k}^2$, and $\sigma_{k'}^2$ denote the power allocation variables and the background noise powers for GU $k$ and GU $k'$ at time slot $n$, respectively.
Note that when $k=k'$, $s_{k,k}[n]=1$ models the case of TDMA where only GU $k$ is served at time slot $n$ and the achievable rate can be given by $C_{k,k}^{\mathrm{II,SIC}}[n]$ in \eqref{OMA}.
However, due to the existence of randomly scattered components in Rician fading channels, an outage event occurs when the transmission rate exceeds the achievable data rate.
To capture the potential outage events, we define the effective rate allocation for GU $k$ and GU $k'$ at time slot $n$ as $r_k[n]$ and $r_{k'}[n]$, respectively.
When $s_{k,k'}[n]=1$, $r_k[n]$ can be achieved when the two stages of SIC decoding at GU $k$ are successful, i.e., $ r_{k'}[n] \le C_{k,k'}^{\mathrm{I,SIC}}[n]$ and $r_k[n] \le C_{k,k'}^{\mathrm{II,SIC}}[n]$.
Meanwhile, $r_{k'}[n]$ can be achieved when the direct decoding at GU $k'$ is successful, i.e., $r_{k'}[n] \le C_{k,k'}^{\mathrm{NSIC}}[n]$.
Besides, when $s_{k,k}[n]=1$, $r_k[n]$ can be achieved when $r_{k}[n] \le C_{k,k}^{\mathrm{II,SIC}}[n]$.

Our design aims to satisfy the minimum data rate requirement of each user while taking into account the potential outages of both SIC decoding and direct decoding.
Therefore, we formulate the average transmission rate of user $k$ during the whole flight period as follows:\vspace{-1mm}
\begin{equation}
\bar R_k = \underbrace{\frac{1}{N} \sum_{n=1}^N  \sum_{\substack{k'=1\\k'\neq k}}^K s_{k,k'}[n] r_{k}[n]}_{\text{GU $k$ as a SIC user}} + \underbrace{\frac{1}{N} \sum_{n=1}^N  \sum_{\substack{k'=1\\k'\neq k}}^K s_{k'\!,k}[n] r_{k}[n]}_{\text{GU $k$ as a non-SIC user}} + \underbrace{\frac{1}{N} \sum_{n=1}^N s_{k,k}[n] r_{k}[n]}_{\text{GU $k$ as an OMA user}}, \\[-1mm]
\end{equation}
where the first term denotes the average transmission rate of GU $k$ as a SIC user, the second term denotes the average transmission rate of GU $k$ as a non-SIC user, and the third term represents the average transmission rate of GU $k$ as an OMA user.

\vspace{-3mm}
\subsection{Power Consumption Model}
\vspace{-1mm}

The power consumption of the UAV plays an important role in UAV-based communications due to its small-size onboard battery with limited energy capacity.
The system power consumption consists of the UAV's communication power and the flight power.
The communication power of the UAV at time slot $n$ can be given by\vspace{-1.5mm}
\begin{equation}
P_{\mathrm{comm}}[n] = \sum_{k=1}^K  \sum_{k'\neq k}^K  s_{k,k'}[n] \left(p_k[n] \!+\! p_{k'}[n]\right) + \sum_{k=1}^K s_{k,k}[n] p_k[n].\\[-1.5mm]
\end{equation}
Note that the first term denotes the communication power consumption for NOMA users and the second term represents the communication power of the users selected to operate in OMA mode.
In this system, we consider a rotary wing UAV as it has a higher maneuverability than fixed wing UAVs.
According to \cite{rotary_wing_power,DWK_2018_solar_power_comm}, the flight power consumption of a rotary wing UAV at time slot $n$ is given by\vspace{-1mm}
\begin{equation}\label{flight_power}
\hspace{-4mm}P_{\mathrm{fly}}[n]\!=\! \underset{\mathrm{Bladeprofile \, power}}{\underbrace{P_o \! \left(\! 1 \!+\! \frac{3 (v_x^2[n] \hspace{-1mm}+\hspace{-1mm} v_y^2[n]) }{\Omega^2 r^2} \!\right)}} \!+ \underset{\mathrm{Induced \, power}}{\underbrace{\frac{P_i v_0}{v_x^2[n] \hspace{-1mm}+\hspace{-1mm} v_y^2[n]}}}
+ \underset{\mathrm{Parasite \, power}}{\underbrace{\frac{1}{2} d_0 \rho s A_{\mathrm r} (v_x^2[n] \!+\! v_y^2[n])^{3/2}}} \!+\hspace{-4.5mm} \underset{\mathrm{\hspace{1mm} Vertical \, flight \, power}}{\underbrace{G v_{\mathrm{z}}[n]}} \hspace{-7.5mm}, \hspace{-6mm}\\[-1.5mm]
\end{equation}
where the velocity of the UAV in 3D Cartesian coordinate is denoted as $\mathbf{v}[n] = [v_x[n],v_y[n],$ $v_z[n]]^{\mathrm T} \in \mathbb{R}^{3 \times 1}$.
The physical meanings of the parameters in \eqref{flight_power} are summarized in Table \ref{notations}.
In \eqref{flight_power}, the first three components are related to the horizontal flight power and the last component, representing the vertical flight power consumption, plays an important role in controlling the UAV's flight altitude.
In particular, it is expected that optimizing the vertical velocity, $v_{\mathrm z}[n]$, can affect the flight endurance and the flight power consumption.

\linespread{1.05}
\begin{table}[t] \vspace*{-2mm}
	\centering
	\scriptsize
	\caption{Physical meaning of parameters in flight power consumption model of UAV \cite{rotary_wing_power}.} \label{notations}
	\vspace*{-4mm}
	\begin{tabular}{ c | c | c }
		\hline		
		Parameters        & Physical meaning                          & Simulation values \\ \hline
        $G$               & Weight of UAV                             & 20 (Newton) \\
		$\Omega$          & Blade angular velocity       			  & 300 (radians/second)\\
		$r$               & Rotor radius                              & 0.4 (meter)\\
		$\rho$            & Air density                               & 1.225 ($\mathrm{kg/m^3}$) \\
		$s$               & Rotor solidity                            & 0.05 ($\mathrm{m^3}$) \\
		$A_{\mathrm r}$   & Rotor disc area                           & 0.503 ($\mathrm{m^2}$) \\
		$v_0$             & Induced velocity for rotor in forwarding flight & 4.03 (meter/second) \\
		$d_0$             & Fuselage drag ratio                       & 0.3 \\
        $P_o$             & Blade profile power in hovering status    & 79.86 (watt) \\
		$P_i$             & Induced power in hovering status          & 88.63 (watt) \\
		\hline
	\end{tabular}
	\vspace{-6mm}
\end{table}

On the other hand, in practice, the IRS is usually mounted on the building exterior which is accessible to energy source.
Besides, the IRS is nearly passive and its operation power is a constant which is much lower than that of the communication and flight power consumption of the UAV \cite{8811733,rotary_wing_power}.
Therefore, we ignore the IRS power consumption in the considered system.

\vspace{-2mm}
\section{Problem Formulation}
\vspace{-1mm}

The optimization problem for minimizing the average total power consumption via jointly designing the user scheduling $\mathcal{S} = \{s_{k,k'}[n],\forall n,k,k'\}$,
the power allocation $\mathcal{P} = \{p_k[n],$ $\forall n,k\}$, the effective transmission rate $\mathcal{R} = \{r_k[n], \forall n,k\}$\footnote{Note that the transmission rate $r_k[n]$ is optimized to satisfy the outage probability constraints based on the practical altitude-dependent Rician fading channel model while only the statistical CSI is available for resource allocation.},
the UAV's 3D trajectory $\mathcal{T} = \{\mathbf{t}[n], \forall n\}$,
the UAV's 3D flight velocity $\mathcal{V} = \{\mathbf{v}[n], \forall n\}$,
and the phase control policy of the IRS $\bm \Phi = \{\phi_{m_{\mathrm Rx},m_{\mathrm Ry}}[n],$ $\forall n,m_{\mathrm Rx},m_{\mathrm Ry}\}$ is formulated as:\vspace{-1mm}
\begin{align} & \hspace{-12mm} \underset{\mathcal{S},\mathcal{P},\mathcal{R},\mathcal{T},\mathcal{V}, \bm \Phi}{\mathrm{minimize}}\,\, \frac{1}{N} \sum_{n=1}^N P_{\mathrm{comm}}[n] + \frac{1}{N} \sum_{n=1}^N P_{\mathrm{fly}}[n] \label{proposed_formulation_origion} \\[-1.5mm]
\mathrm{s.t.} \, \mathrm{C1}&\!: s_{k,k'}[n] \in \{0,1\}, \forall n,k,k',
\hspace{0.8mm}  \mathrm{C2}: \sum_{k=1}^K  \sum_{k'=1}^K  s_{k,k'}[n] \leq 1, \forall n, \notag \\[-1.5mm]
\mathrm{C3}&\!: p_k[n] \geq 0, \forall n,k,
\hspace{17mm} \mathrm{C4}: P_{\mathrm{comm}}[n] \leq P_{\mathrm{peak}}, \forall n,
\hspace{15mm} \mathrm{C5}: \bar R_k \geq R_{\min_{k}}, \forall k, \notag \\
\mathrm{C6}&\!: \text{Pr} \Big(\! s_{k,k'}[n] r_{k'}[n] \hspace{-1mm}\leq\hspace{-1mm} s_{k,k'}[n] C_{k,k'}^{\mathrm{I,SIC}}[n], s_{k,k'}[n] r_k[n] \hspace{-1mm}\leq\hspace{-1mm} s_{k,k'}[n] C_{k,k'}^{\mathrm{II,SIC}}[n] \!\Big) \hspace{-1mm}\geq\hspace{-1mm} 1 \hspace{-1mm}-\! \varepsilon_k^{\mathrm{SIC}} \!, \! \forall n,k' \!\!\neq\! k, \notag \\[-0.5mm]
\mathrm{C7}&\!: \text{Pr} \Big( s_{k'\!,k}[n] r_k[n] \leq s_{k'\!,k}[n] C_{k'\!,k}^{\mathrm{NSIC}}[n] \Big) \geq 1-\varepsilon_k^{\mathrm{NSIC}}, \! \forall n,k'\!\neq\! k, \notag \\[-0.5mm]
\mathrm{C8}&\!: \text{Pr} \Big( s_{k,k}[n] r_k[n] \leq s_{k,k}[n] C_{k,k}^{\mathrm{II,SIC}}[n] \Big) \geq 1-\varepsilon_k^{\mathrm{OMA}}, \! \forall n,k, \notag \\[-1mm]
\mathrm{C9}&\!: \mathbf{t}[n+1] = \mathbf{t}[n] + \mathbf{v}[n] \tau, n=1,\ldots,N-1, \notag \\[-0.5mm]
\mathrm{C10}&\!: \mathbf{t}[1] = \mathbf{t}_0,
\hspace{26.6mm} \mathrm{C11}: \mathbf{t}[N] = \mathbf{t}_{\mathrm F}, \notag \\[-0.5mm]
\mathrm{C12}&\!: \mathbf{t}_{\min} \leq \mathbf{t}[n] \leq \mathbf{t}_{\max}, \forall n,
\hspace{2.8mm} \mathrm{C13}: \|\mathbf{v}[n+1]-\mathbf{v}[n]\| \leq V_{\mathrm{acc}} \tau, \forall n, \notag \\[-0.5mm]
\mathrm{C14}&\!: \|\mathbf{v}[n]\| \leq V_{\max}, \forall n,
\hspace{9.8mm} \mathrm{C15}: 0 \leq \phi_{m_{\mathrm Rx},m_{\mathrm Ry}}[n] < 2 \pi, \forall n,m_{\mathrm Rx},m_{\mathrm Ry}. \notag \\[-11mm]\notag
\end{align}
Note that $\mathrm{C1}$ defines the user scheduling variable and $\mathrm{C2}$ guarantees that at most two users are scheduled at each time slot.
$\mathrm{C3}$ is the non-negative constraint for the transmit power from the UAV to GU $k$ and $P_{\mathrm{peak}}$ in $\mathrm{C4}$ represents the peak transmission power of the UAV at each time slot.
Constraint $\mathrm{C5}$ is introduced to limit the minimum data rate for each user.
Constraints $\mathrm{C6}-\mathrm{C8}$ represent the maximum outage\footnote{Note that the channel outage happens when the effective transmission rate is larger than the channel capacity.} probability constraints for the SIC decoding at SIC users, the direct decoding at non-SIC users\footnote{In this system, we assume that the SIC user did not tempt to decode its own message if the first stage is failed, as commonly adopted in \cite{Zhiqiang_DC,8529214}.}, and the direct decoding at OMA users, respectively, where $\varepsilon_k^{\mathrm{SIC}}>0$, $\varepsilon_k^{\mathrm{NSIC}}>0$, and $\varepsilon_k^{\mathrm{OMA}}>0$ are corresponding maximum tolerable outage probabilities.
Constraint $\mathrm{C9}$ denotes the relationship between the UAV's 3D trajectory and its flight velocity\footnote{Note that the UAV's flight velocity is a function of its trajectory for a given time slot duration $\tau$.
However, directly expressing the total power consumption in terms of UAV's trajectory or expressing the effective channel gain in terms of UAV's flight velocity would lead to an intractable formulation.
Thus, we introduce variables of the UAV's trajectory and flight velocity to simplify the problem at hand.}.
$\mathrm{C10}$ and $\mathrm{C11}$ denote the starting and the final locations of the UAV, respectively.
Parameters $\mathbf{t}_{\min}$ and $\mathbf{t}_{\max}$ in constraint $\mathrm{C12}$ limit the maximum service area of the UAV.
$V_{\mathrm{acc}}$ and $V_{\max}$ in constraints $\mathrm{C13}$ and $\mathrm{C14}$ denote the maximum flight acceleration and the maximum flight velocity, respectively.
$\mathrm{C15}$ limits the range of phase control of the IRS.

In practice, although the transmit power consumption is much lower than the flight power consumption \cite{rotary_wing_power}, minimizing both $p_{\mathrm{comm}}[n]$ and $P_{\mathrm{fly}}[n]$ in the objective function is necessary since minimizing $p_{\mathrm{comm}}[n]$ can limit the induced interference level imposed to conventional terrestrial cellular networks.



\vspace{-3mm}
\section{Problem Solution}
\vspace{-2mm}

The formulated problem in \eqref{proposed_formulation_origion} is a non-convex optimization problem and there is no systematic and efficient method to obtain the globally optimal solution.
In the following, we first simplify the studied problem by exploiting its special structure at the optimality.
Then, a computationally-efficient suboptimal algorithm is proposed to obtain a high-quality solution.

\vspace{-3mm}
\subsection{Phase Control and Outage-guaranteed Effective Channel Gain}
\vspace{-1.5mm}

Since the randomly scattered components of the channels, e.g. $\Delta h_k^{\mathrm{AG}}[n]$ and $\Delta \mathbf{h}_k^{\mathrm{RG}}[n]$, are unknown, optimizing the phase control policy of the IRS has to rely on the LoS components of the channels, $\mathbf{h}^{\mathrm{AR}}[n]$, $h_k^{\mathrm{AG,LoS}}[n]$, and $\mathbf{h}_k^{\mathrm{RG,LoS}}$.
{To facilitate our design, we propose a efficient closed-form suboptimal phase control policy and derive the effective channel gain based on the distributions of $\Delta h_k^{\mathrm{AG}}[n]$ and $\Delta \mathbf{h}_k^{\mathrm{RG}}[n]$.
Note that the proposed design can significantly reduce the required signaling overhead for CSI acquisition and phase control.

To determine the IRS phase control, since GU $k$ is designed to perform SIC decoding and it is more likely to suffer from the channel outage than GU $k'$, we assume that the IRS is always controlled to coherently combine the LoS channels of GU $k$ when $s_{k,k'}[n] = 1$.
In the following, we summarize a suboptimal phase control policy in a theorem.
\vspace{-1mm}
\begin{Thm} \label{optimize_phase_shift}
A suboptimal phase control policy of the IRS at time slot $n$ $\phi_{m_{\mathrm Rx},m_{\mathrm Ry}}[n]$ for minimizing the total system power consumption is given by\vspace{-1mm}
\begin{align}\label{phase_shift}
\!& \phi_{m_{\mathrm Rx},m_{\mathrm Ry}}[n] = \sum_{k=1}^K \!\sum_{k'=1}^K s_{k,k'}[n] \! \left( \frac{2 \pi \Delta_{\mathrm Rx}}{\lambda_{\mathrm c}} (m_{\mathrm Rx}-1) \left( \sin \theta_{k}^{\mathrm{RG}} \cos \xi_{k}^{\mathrm{RG}} - \sin \theta^{\mathrm{RA}}[n] \cos \xi^{\mathrm{RA}}[n] \right) \right.\notag \\[-0.5mm]
\!& \!\!\left.+ \frac{2 \pi \Delta_{\mathrm Ry}}{\lambda_{\mathrm c}} (m_{\mathrm Ry} \!-\! 1)\!\! \left( \sin \theta_{k}^{\mathrm{RG}} \sin \xi_{k}^{\mathrm{RG}} \!\!-\! \sin \theta^{\mathrm{RA}}[n] \sin \xi^{\mathrm{RA}}[n] \right) \!\!+\! \frac{2 \pi}{\lambda_{\mathrm c}} \!\!\left( d_k^{\mathrm{AR}}[n] \!+\! d_k^{\mathrm{RG}} \!\!-\! d_k^{\mathrm{AG}}[n]\right)\!\!\right)\!\!. \hspace{-2mm}\\[-10mm]\notag
\end{align}
\end{Thm}\vspace{-1mm}
\emph{\quad Proof: Please refer to the appendix for a proof of Theorem \ref{optimize_phase_shift}. \qed
}

Applying \eqref{phase_shift} to \eqref{h_k_AG}--\eqref{h_k_RG_LoS} yields the effective channels from the UAV to GU $k$ and GU $k'$ at time slot $n$ are given by\vspace{-1mm}
\begin{align}
h_{k}[n]=& \sqrt{\frac{\beta_0 \kappa_k^{\mathrm{AG}}[n]}{A_k[n]}} + \sqrt{\frac{\beta_0^2 \kappa^{\mathrm{RG}} M_{\mathrm R}^2}{B_k[n]}} + \sqrt{\frac{\beta_0}{A_k[n]}} \Delta h_k^{\mathrm{AG}}[n] + \sqrt{\frac{\beta_0^2 M_{\mathrm R}^2}{B_k[n]} } \Delta h_k^{\mathrm{RG}}[n] \, \text{and} \label{optimal_effective_channel}\\[-0.5mm]
h_{k'}[n]=& \sqrt{\frac{\beta_0 \kappa_{k'}^{\mathrm{AG}}[n]}{A_{k'}[n]}} e^{-j \frac{2\pi}{\lambda}(d_{k'}^{\mathrm{AG}}[n]+d_{k}^{\mathrm{AG}}[n])} + \sqrt{\frac{\beta_0^2 \kappa^{\mathrm{RG}} }{B_{k'}[n]}} e^{-j \frac{2\pi}{\lambda} (d_{k'}^{\mathrm{RG}}-d_{k}^{\mathrm{RG}})} \! \sum_{m_{\mathrm Rx}=1}^{M_{\mathrm Rx}} \sum_{m_{\mathrm Ry}=1}^{M_{\mathrm Ry}} \exp \left(-j\frac{2\pi}{\lambda} \right.\notag \\[-2mm]
&  \left[\Delta_{\mathrm Rx} (m_{\mathrm Rx}\!-1) (\sin \theta_{k'}^{\mathrm{RG}} \! \cos \xi_{k'}^{\mathrm{RG}}-\sin \theta_{k}^{\mathrm{RG}} \! \cos \xi_{k}^{\mathrm{RG}}) \!+\! \Delta_{\mathrm Ry} (m_{\mathrm Ry}\!-1) (\sin \theta_{k'}^{\mathrm{RG}} \! \sin \xi_{k'}^{\mathrm{RG}} \right. \notag \\[-0.5mm]
& \left.- \sin \theta_{k}^{\mathrm{RG}} \! \sin \xi_{k}^{\mathrm{RG}})\right] \bigg) + \sqrt{\frac{\beta_0}{A_{k'}[n]}} \Delta h_{k'}^{\mathrm{AG}}[n] + \sqrt{\frac{\beta_0^2 }{B_{k'}[n]} } \Delta h_{k'}^{\mathrm{RG}}[n], \label{optimal_effective_channel_other_k} \\[-11mm]\notag
\end{align}
respectively, where $A_k[n] = (d_k^{\mathrm{AG}}[n])^{\alpha^{\mathrm{AG}}}(1 + \kappa_k^{\mathrm{AG}}[n])$, $B_k[n] = (d^{\mathrm{AR}}[n])^{\alpha^\mathrm{AR}} (d_k^{\mathrm{RG}})^{\alpha^{\mathrm{RG}}} (1+\kappa^{\mathrm{RG}})$, $A_{k'}[n] = (d_{k'}^{\mathrm{AG}}[n])^{\alpha^{\mathrm{AG}}}(1 + \kappa_{k'}^{\mathrm{AG}}[n])$, and $B_{k'}[n] = (d^{\mathrm{AR}}[n])^{\alpha^\mathrm{AR}} (d_{k'}^{\mathrm{RG}})^{\alpha^{\mathrm{RG}}} (1+\kappa^{\mathrm{RG}})$.
We can observe that $h_k[n]$ and $h_{k'}[n]$ follow the Gaussian distribution with mean, $\sqrt{\frac{\beta_0 \kappa_k^{\mathrm{AG}}[n]}{A_k[n]}} + \sqrt{\frac{\beta_0^2 \kappa^{\mathrm{RG}} M_{\mathrm R}^2}{B_k[n]} }$ and $\sqrt{\frac{\beta_0 \kappa_{k'}^{\mathrm{AG}}[n]}{A_{k'}[n]}} + \sqrt{\frac{\beta_0^2 \kappa^{\mathrm{RG}}}{B_{k'}[n]} }$, as well as variance, $\frac{\beta_0}{A_k[n]} + \frac{\beta_0^2 M_{\mathrm R}^2}{B_k[n]}$ and $\frac{\beta_0}{A_{k'}[n]} + \frac{\beta_0^2 }{B_{k'}[n]}$, respectively.
In other words, the end-to-end effective channels of GUs still follows the altitude-dependent Rician fading with our proposed phase control policy\footnote{The obtained closed-form IRS phase shift in \eqref{phase_shift} is the optimal solution for the case of OMA.}.
Note that the outage probability constraints $\mathrm{C6}-\mathrm{C8}$ of the formulated optimization problem \eqref{proposed_formulation_origion} are active at the optimal point.
Then, constraints $\mathrm{C6}-\mathrm{C8}$ can be rewritten as\footnote{Note that since outage constraints $\mathrm{C6}-\mathrm{C8}$ are inactive if $s_{k,k'}[n]=0$, we only consider the situation of $s_{k,k'}[n]=1$ while handling the intractable constraints $\mathrm{C6}-\mathrm{C8}$ in the following process.}\vspace{-2mm}
\begin{align}
\hspace{-3mm} 1\!-\! \varepsilon_{k}^{\mathrm{SIC}}\! &= \text{Pr} \left( s_{k,k'}[n] r_{k'}[n] \leq s_{k,k'}[n] C_{k,k'}^{\mathrm{I,SIC}}[n] , s_{k,k'}[n] r_{k}[n] \leq s_{k,k'}[n] C_{k,k'}^{\mathrm{II,SIC}}[n]\right) \notag \\[-0.5mm]
\hspace{-3mm}&= \text{Pr} \left( \frac{\sigma_{k}^2 (2^{r_{k'}[n]} - 1)}{p_{k'}[n] - p_{k}[n](2^{r_{k'}[n]} - 1)} \leq |h_{k}[n]|^2, \frac{\sigma_k^2 (2^{r_{k}[n]} -1)}{p_k[n]} \leq |h_{k}[n]|^2 \right) \notag \\[-0.5mm]
\hspace{-3mm}&= \max \left\{ 1 - F_{n,k} \left( \frac{\sigma_{k}^2 (2^{r_{k'}[n]} - 1)}{p_{k'}[n] - p_{k}[n](2^{r_{k'}[n]} - 1)} \right), 1 - F_{n,k} \left(\frac{\sigma_k^2 (2^{r_{k}[n]} - 1)}{p_k[n]}\right) \right\}, \hspace{-1mm} \label{c5outage} \\[-0.5mm]
\hspace{-3mm} 1\!-\! \varepsilon_{k}^{\mathrm{NSIC}}\! &= \text{Pr} \left( s_{k'\!,k}[n] r_{k}[n] \leq s_{k'\!,k}[n] C_{k'\!,k}^{\mathrm{NSIC}}[n] \right) \notag \\[-0.5mm]
\hspace{-3mm}&= \text{Pr} \left( \frac{\sigma_{k}^2 (2^{r_{k}[n]} v-\! 1)}{p_{k}[n] \!-\! p_{k'}[n](2^{r_{k}[n]} \!-\! 1)} \!\leq\! |h_{k}[n]|^2 \!\right) \!=\! 1 \!-\! F_{n,k} \!\left( \frac{\sigma_{k}^2 (2^{r_{k}[n]} \!-\! 1)}{p_{k}[n] \!-\! p_{k'}[n](2^{r_{k}[n]} \!-\! 1)} \right)\hspace{-1.2mm}, \hspace{-2mm} \label{c6outage}\\[-0.5mm]
\hspace{-3mm} 1\!-\! \varepsilon_{k}^{\mathrm{OMA}}\! &= \text{Pr} \left( s_{k,k}[n] r_{k}[n] \leq s_{k,k}[n] C_{k,k}^{\mathrm{II,SIC}}[n] \right) \notag \\[-0.5mm]
\hspace{-3mm}&= \text{Pr} \left( \frac{\sigma_k^2 (2^{r_{k}[n]} -1)}{p_k[n]} \leq |h_{k}[n]|^2 \right) = 1 - F_{n,k} \left(\frac{\sigma_k^2 (2^{r_{k}[n]} -1)}{p_k[n]}\right), \label{c7outage} \\[-12mm]\notag
\end{align}
respectively.
Note that $F_{n,k}(\cdot)$ is the cumulative distribution function (CDF) of the random variable $|h_{k}[n]|^2$, which is given by \cite{5506181,3d_trajectory}\vspace{-0.5mm}
\begin{align}
\hspace{-4mm} F_{n,k}(\chi)&= 1- Q_1 \!\left( \sqrt{ \frac{\beta_0 \kappa_k^{\mathrm{AG}}[n] B_k[n]}{D_k[n]}} \!+\! \sqrt{ \frac{\beta_0^2 \kappa^{\mathrm{RG}} M_{\mathrm R}^2 A_k[n]}{D_k[n]}} , \sqrt{\frac{A_k[n]  B_k[n]}{D_k[n]}} \chi\right) \text{and}\hspace{-4mm} \label{marcum_q_function} \\[-0.5mm]
\hspace{-4mm} F_{n,k}(f_{k}^{\mathrm{NSIC}}[n]) &=  1- Q_1 \!\left( \sqrt{ \frac{\beta_0 \kappa_{k}^{\mathrm{AG}}[n] B_{k}[n]}{D_{k}[n]}} \!+\! \sqrt{\frac{\beta_0^2 \kappa^{\mathrm{RG}} A_{k}[n]}{D_{k}[n]}} ,\sqrt{ \frac{A_{k}[n] B_{k}[n]}{D_{k}[n]}} f_{k}^{\mathrm{NSIC}}[n] \right)\hspace{-1mm},  \hspace{-4mm}  \label{marcum_q_function_3} 
\end{align}
where $\chi \in\{f_k^{\mathrm{I,SIC}}[n], f_k^{\mathrm{II,SIC}}[n], f_k^{\mathrm{OMA}}[n]\}$, $f_k^{\mathrm{SIC}}[n] = \min \left\{f_{k}^{\mathrm{I,SIC}}[n], f_{k}^{\mathrm{II,SIC}}[n] \right\}$, and $D_k[n] = \beta_0 B_k[n] \!+\! \beta_0^2 M_{\mathrm R}^2 A_k[n]$.
Note that $f_k^{\mathrm{SIC}}[n]$, $f_{k}^{\mathrm{NSIC}}[n]$, and $f_k^{\mathrm{OMA}}[n]$ denote the outage-guaranteed effective channel gain for GU $k$ when it is selected as a SIC user, a non-SIC user, and an OMA user, respectively.
Function $Q_1(a,b)$ is the standard Marcum-Q function \cite{1055327}.
In general, there is no closed-form expression for \eqref{marcum_q_function} and \eqref{marcum_q_function_3}.
More importantly, their inverse functions, i.e., $F_{n,k}^{-1}(f_{k}^{\mathrm{I,SIC}}[n])$, $F_{n,k}^{-1}(f_{k}^{\mathrm{II,SIC}}[n])$, $F_{n,k}^{-1}(f_{k}^{\mathrm{NSIC}}[n])$, and $F_{n,k}^{-1}(f_{k}^{\mathrm{OMA}}[n])$, for returning outage-guaranteed effective channel gains, $f_{k}^{\mathrm{I,SIC}}[n]$, $f_{k}^{\mathrm{II,SIC}}[n]$, $f_{k}^{\mathrm{NSIC}}[n]$, and $f_{k}^{\mathrm{OMA}}[n]$ are intractable functions with respect to (w.r.t.) the 3D trajectory of the UAV, $\mathbf{t}[n]$.
On the other hand, although the value of the Marcum-Q function can be found via a lookup table, it does not facilitate the overall resource allocation design.

\begin{figure}[t]
  \vspace*{-4mm}
  \centering
  \includegraphics[width=3.9 in]{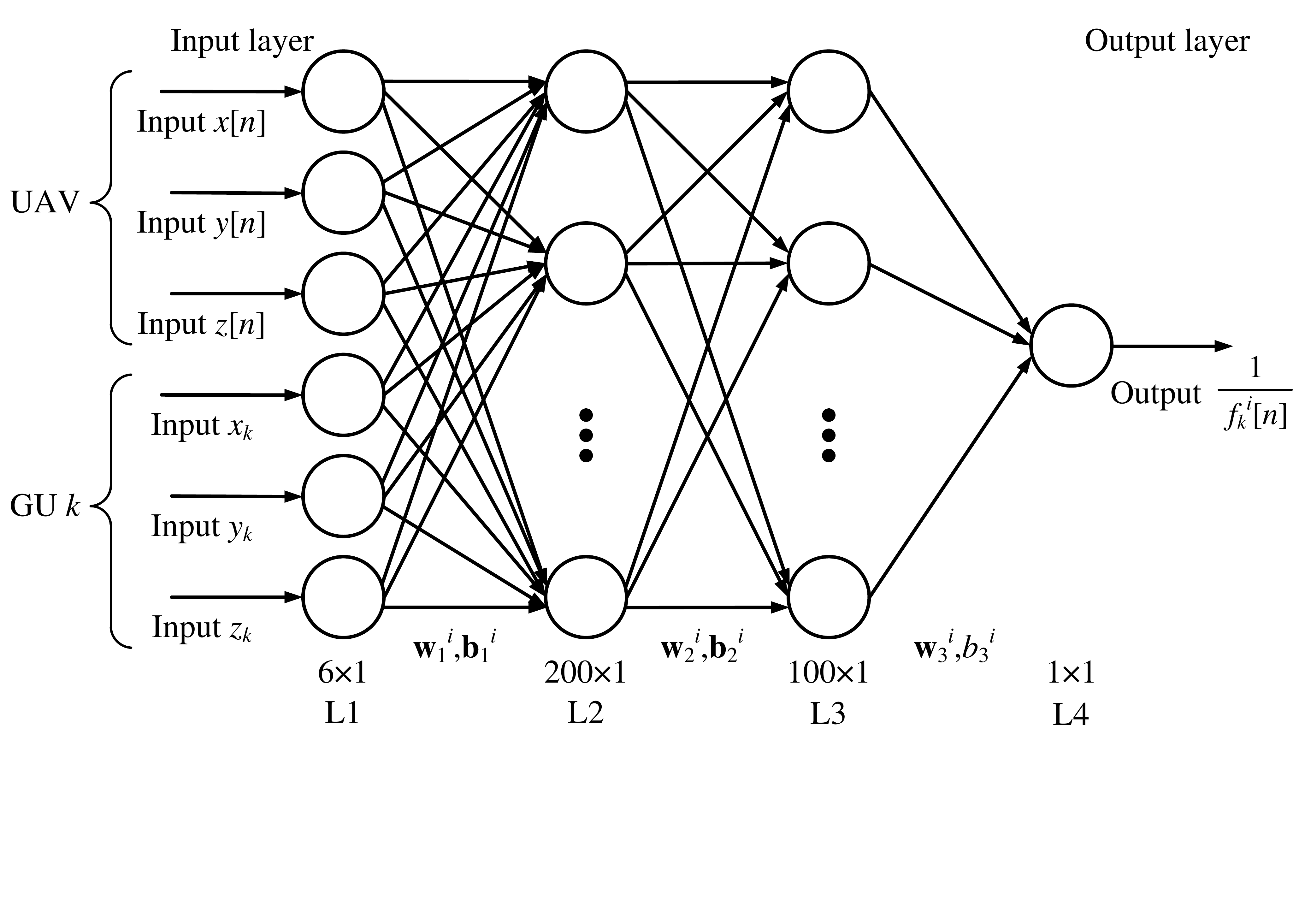}
  \vspace*{-17mm}
  \caption{The structure of the employed three-layer feedforward neural network.}
  \label{neural_network}
  \vspace*{-7mm}
\end{figure}

To overcome the intractability in \eqref{marcum_q_function} and \eqref{marcum_q_function_3}, also, to strike a balance between the system performance and the computational complexity, in this paper, we adopt a DNN approach\footnote{Note that although existing data regression methods, e.g. discriminant analysis and stochastic modeling, can be adopted to approximate the sophisticated effective channel gains \cite{kaya2009characterizing,fan2013two}, they either incur high computational complexity or result in limited performance.
Besides, their intractability do not facilitate the design of computationally efficient resource allocation policy.} \cite{liu2019deep,liu2020deep} to approximate the outage-guaranteed effective channel gain $f_k^{i}[n]$ for different schemes $i \in \{\mathrm{SIC},\mathrm{NSIC},\mathrm{OMA}\}$ as a tractable function w.r.t. the 3D trajectory of the UAV and location of GUs.
Fig. \ref{neural_network} shows the structure of a three-layer feedforward neural network \cite{svozil1997introduction}.
{Thus, for each location of the UAV, $\mathbf{t}[n]$, the location of GU $k$, $\mathbf{l}_k$, and a given outage probability, $\varepsilon_k^{i}$ as the generated data sets for the DNN approach, we can generate the numerical data of $\frac{1}{f_k^{i}[n]}$ based on \eqref{kappa_k_AG}--\eqref{h_k_RG_LoS}, \eqref{c5outage}--\eqref{marcum_q_function_3} which serve as labels for neural network training\footnote{The reason for adopting $\frac{1}{f_k^{i}[n]}$ as a label rather than $f_k^{i}[n]$ is that the former can be interpreted as the path loss between the UAV and the desired GU's location which directly depends on the UAV's trajectory.
{Besides, the training and testing data sets are prepared by the mentioned generated data sets that adopt $70$\% and $30$\% of the data sets, respectively.}}.}
After offline training based on the generated samplings, we can then obtain a well-trained neural network.
That is, for given maximum tolerable outage probabilities $\varepsilon_{k}^{i}$, we obtain the approximated outage-guaranteed user location-aware effective channel gain for GU $k$ as 
\begin{equation}\label{neural_network_model}
f_{k}^{i}[n] \approx \frac{1}{(\mathbf{w}_{3}^{i})^{\mathrm H} \left[ \mathbf{w}_{2}^{i} \left[\mathbf{w}_{1}^{i} \mathbf{q}_k[n] + \mathbf{b}_{1}^{i} \right]^+ + \mathbf{b}^{i}_{2}\right]^+ + b_3^i}. 
\end{equation}
Note that $\left[\mathbf{w}_{1}^{i} \mathbf{q}_k[n] + \mathbf{b}_{1}^{i} \right]^+$ is the rectified linear unit (ReLU) function for GU $k$ adopting scheme $i$, which is a convex function w.r.t. $\mathbf{q}_k[n]$.
Vector $\mathbf{q}_k[n] = [\mathbf{t}[n];\mathbf{l}_k] \in \mathbb{R}^{6 \times 1}$ collects the trajectory of the UAV and the location of GU $k$ at time slot $n$.
Parameters $\mathbf{w}_{1}^{i} \in \mathbb{R}^{200\times6}$, $\mathbf{b}_{1}^{i} \in \mathbb{R}^{200\times1}$, $\mathbf{w}_{2}^{i} \in \mathbb{R}^{100\times200}$, $\mathbf{b}_{2}^{i} \in \mathbb{R}^{100\times1}$, $\mathbf{w}_{3}^{i} \in \mathbb{R}^{100\times1}$, and $b_{3}^{i} \in \mathbb{R}$ are the well-trained weights and biases for scheme $i$ between layer $1$ and layer $2$, layer $2$ and layer $3$, as well as layer $3$ and layer $4$, respectively, as shown in Fig. \ref{neural_network}.
To verify the approximation accuracy, as shown in Fig. \ref{comparisons_xy} and Fig. \ref{comparisons_z}, we present the numerical result of $f_k^{i}[n]$ based on \eqref{marcum_q_function} and \eqref{marcum_q_function_3} as well as the approximated value by the neural network model according to \eqref{neural_network_model} for different dimensions, respectively.
We can observe from them that the numerical result of $f_k^{i}[n]$ based on \eqref{kappa_k_AG}--\eqref{h_k_RG_LoS}, \eqref{c5outage}--\eqref{marcum_q_function_3} closely match their predicted effective channel gains obtained by our well-trained numerical network model.
{Besides, with a sufficient number of training data the normalized mean square error (NMSE) between the numerical channel gain and the outage-guaranteed effective channel gain via the DNN approach is less than $0.005$ in our considered setting that is negligible for resource allocation design \cite{hogg2005introduction,tugnait1997identification}.}
In particular, there is a non-trivial trade-off between the flight altitude, outage-guaranteed effective channel gain, and the horizontal distance between the UAV and the GUs.
For example, as shown in Fig. \ref{comparisons_z}, when the horizontal distance between the UAV and the desired users (e.g. GU $1$ and GU $2$) is large, increasing the flight altitude to a certain extent would increase the effective channel gain.
Specifically, a higher flight altitude can reduce the variance of Rician fading channel in \eqref{h_k_AG} which facilitates a more power-efficient UAV communication.
However, an exceedingly high flight altitude would cause the decrease in outage-guaranteed effective channel gain, as the increased path loss outweights the gain brought by reduced channel uncertainty.
On the other hand, when the horizontal distance between the UAV and the user (e.g. GU $3$) is short, increasing the flight altitude is not beneficial to the effective channel gain since the increased path loss is dominated.
As a result, we set the outage-guaranteed transmission rate for GU $k$ as a SIC user, a non-SIC user, and an OMA user\footnote{Note that we will verify the accuracy of the outage probability obtained by using \eqref{neural_network_model} as the outage-guaranteed effective channel gain in the simulation section.} are given by\vspace{-0.5mm}
\begin{align}
r_{k}[n] &= \log_2 \left( 1 + \frac{p_k[n] f_{k}^{\mathrm{SIC}}[n]}{\sigma_k^2} \!\right), \forall n, k\neq k', \,\text{if}\, s_{k,k'}[n]=1, \label{r_kn} \\[-0.3mm]
r_{k}[n] &= \log_2 \left( 1 + \frac{p_{k}[n] f_{k}^{\mathrm{NSIC}}[n]}{p_{k'}[n] f_{k}^{\mathrm{NSIC}}[n] + \sigma_{k}^2} \!\right), \forall n, k\neq k', \,\text{if}\, s_{k'\!,k}[n]=1, \label{r_kn_2} \,\,\text{and} \\[-0.3mm]
r_{k}[n] &= \log_2 \left( 1 + \frac{p_k[n] f_{k}^{\mathrm{OMA}}[n]}{\sigma_k^2} \right), \forall n,k, \,\text{if}\, s_{k,k}[n]=1, \\[-11mm]\notag
\end{align}
\begin{figure}[t]\vspace*{-2mm}
  \centering
  \begin{minipage}[t]{0.49\textwidth}
      \centering
      \hspace*{-3mm} \includegraphics[width=3.1 in]{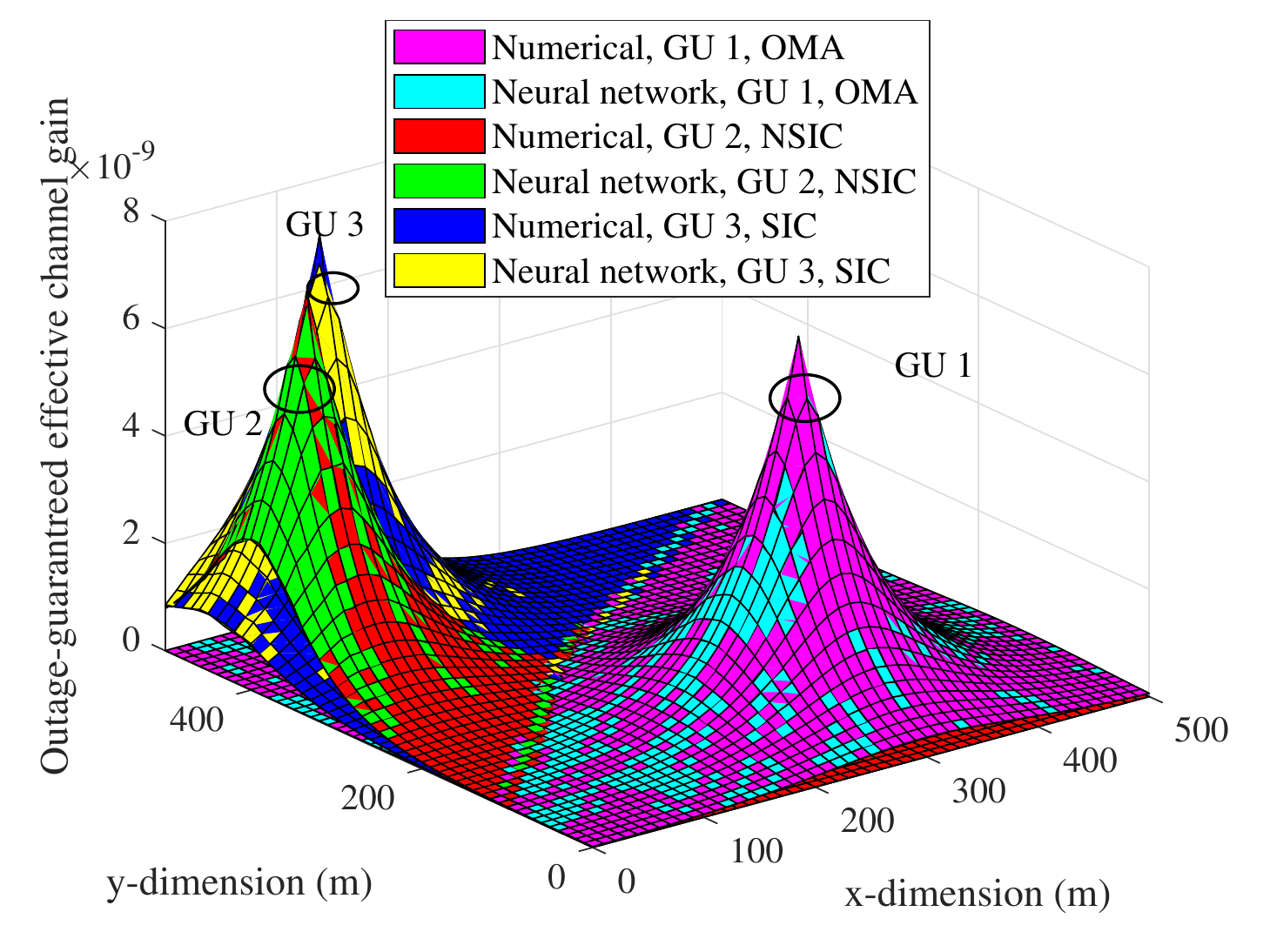} \vspace*{-9mm}
      \caption{Comparisons between the numerical data and the neural network model versus the UAV's location at the x-dimension and the y-dimension for a specific altitude, i.e., $140$ m, when the maximum tolerable outage probability $=$ $0.01$ and $\kappa_{\min}=0$ dB. The location of GUs and the IRS are listed in Table \ref{simulation_setting}.}
      \label{comparisons_xy}
  \end{minipage}\,\,
  \begin{minipage}[t]{0.49\textwidth}
      \centering
      \hspace*{-3mm}\includegraphics[width=3.1 in]{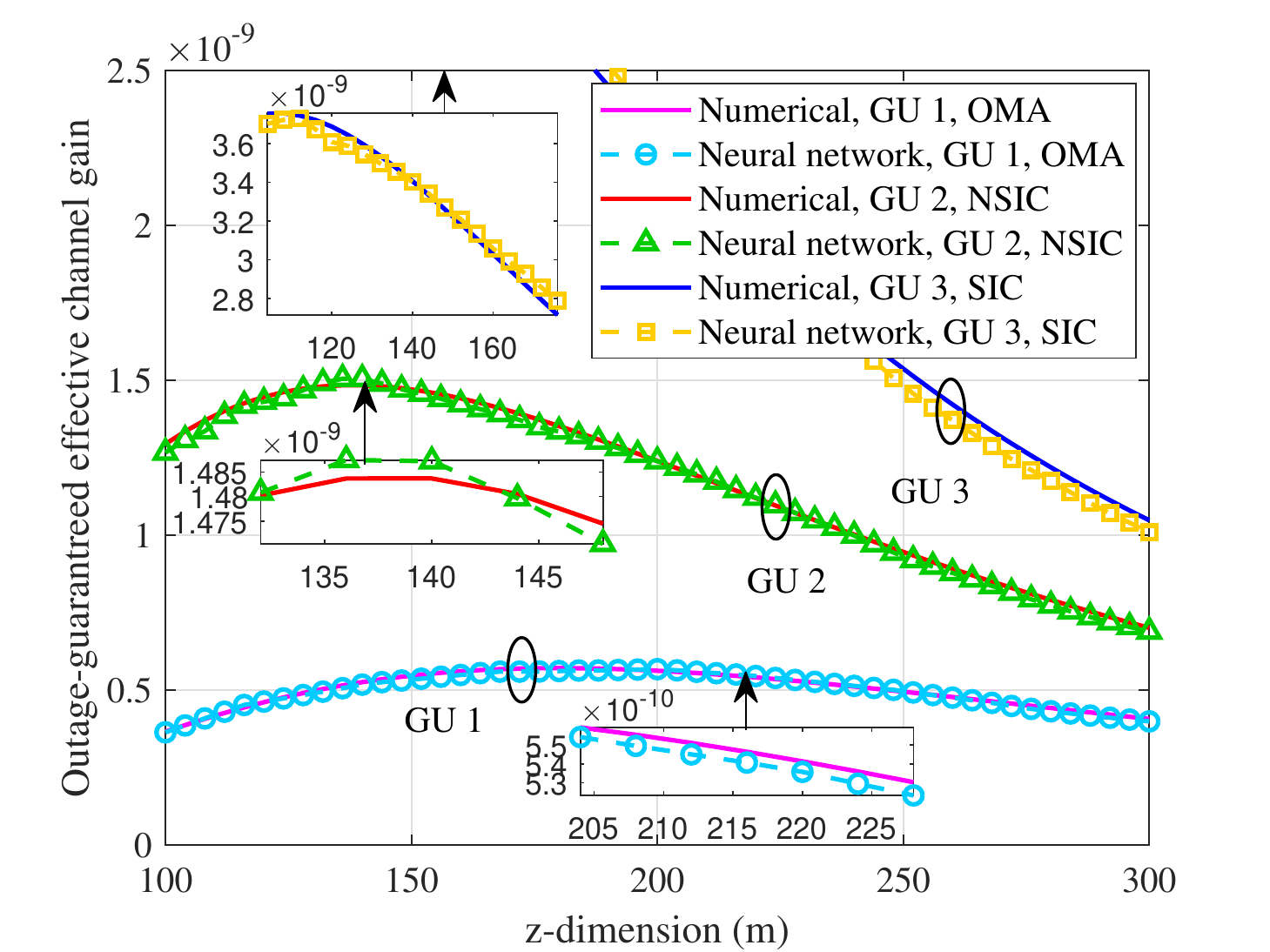} \vspace*{-3.8mm}
      \caption{Comparisons between the numerical data and the neural network model versus the UAV's vertical location at z-dimension with horizontal  locations, i.e., $(170,200)$, $(180,300)$, and $(60,400)$, calculated for GU $1$, GU $2$, and GU $3$, respectively, when the maximum tolerable outage probability $=$ $0.01$ and $\kappa_{\min}=0$ dB.}
      \label{comparisons_z}
  \end{minipage}
  \vspace*{-9mm}
\end{figure}
respectively.
Then, by applying \eqref{r_kn} and \eqref{r_kn_2} to constraint $\mathrm{C5a}$, we can readily reformulate the original optimization formulation in \eqref{proposed_formulation_origion} as the following problem:\vspace{-1mm}
\begin{align}
& \underset{\mathcal{S},\mathcal{P},\mathcal{T},\mathcal{V},\mathcal{Q}} {\mathrm{minimize}} \,\,\frac{1}{N} \sum_{n=1}^N P_{\mathrm{comm}}[n] + \frac{1}{N} \sum_{n=1}^N P_{\mathrm{fly}}[n] \label{proposed_formulation_approximated} \\[-1mm]
& \mathrm{s.t.} \,\,\mathrm{C1}-\mathrm{C4}, \mathrm{C9}-\mathrm{C14}, \notag \\[-1mm]
& \overline{\mathrm{C5}}: \,\,\frac{1}{N}  \sum_{n=1}^N \Bigg[ \sum_{\substack{k'=1\\ k'\neq k}}^K \left( s_{k,k'}[n] \log_2 \!\left(\! 1 \!+\! \frac{p_k[n] f_k^{\mathrm{SIC}}[n]}{\sigma_k^2} \right) \!+\! s_{k'\!,k}[n] \log_2 \!\left(\! 1 \!+\! \frac{p_{k}[n] f_{k}^{\mathrm{NSIC}}[n]}{p_{k'}[n] f_{k}^{\mathrm{NSIC}}[n] \!+\! \sigma_k^2} \right)\!\right) \notag \\[-1mm]
& \hspace{8mm}+ s_{k,k}[n] \log_2 \left( 1 + \frac{p_k[n] f_k^{\mathrm{OMA}}[n]}{\sigma_k^2} \right) \Bigg] \geq R_{\min_{k}}, \forall k, \,\,\,\,\,\, \mathrm{C16}:\,\,\mathbf{q}_k[n] = [\mathbf{t}[n];\mathbf{l}_k], \forall n,k, \notag\\[-11mm]\notag
\end{align}
where $\mathcal{Q} = \{\mathbf{q}_k[n], \forall n,k\}$.
Note that we can rewrite $\overline{\mathrm{C5}}$ as\vspace{-0.5mm}
\begin{equation}
\overline{\overline{\mathrm{C5}}}: \frac{1}{N}  \sum_{n=1}^N \Bigg[ \sum_{\substack{k'=1\\ k'\neq k}}^K R_{k,k'}^{\mathrm{SIC}}[n] + \sum_{\substack{k'=1\\ k'\neq k}}^K \left(R_{k'\!,k}^{\mathrm{I}}[n] - R_{k'\!,k}^{\mathrm{II}}[n]\right) + R_{k,k}^{\mathrm{OMA}}[n] \Bigg] \geq R_{\min_{k}}, \forall k,\\[-1mm]
\end{equation}
where\vspace{-0.5mm}
\begin{align}
R_{k,k'}^{\mathrm{SIC}}[n] &= s_{k,k'}[n]\log_2 \left( 1 + \frac{p_k[n] f_k^{\mathrm{SIC}}[n]}{\sigma_k^2} \right),\\[-0.5mm]
R_{k'\!,k}^{\mathrm{I}}[n] &= s_{k'\!,k}[n] \log_2 \!\left( (p_{k}[n] + p_{k'}[n]) f_{k}^{\mathrm{NSIC}}[n] + \sigma_k^2 \right),\\[-0.5mm]
R_{k'\!,k}^{\mathrm{II}}[n] &= s_{k'\!,k}[n] \log_2 \!\left( p_{k'}[n] f_{k}^{\mathrm{NSIC}}[n] + \sigma_k^2 \right)\!, \, \text{and} \hspace{-3mm}\\[-0.5mm]
R_{k,k}^{\mathrm{OMA}}[n] &= s_{k,k}[n]\log_2 \left( 1 + \frac{p_k[n] f_k^{\mathrm{OMA}}[n]}{\sigma_k^2} \right).\\[-11mm]\notag
\end{align}

Although the reformulated problem in \eqref{proposed_formulation_approximated} is more tractable, it is still non-convex due to the coupling between the communication resource allocation variables and the UAV's trajectory design variables.
Now, to obtain an efficient suboptimal solution, we adopt the alternating optimization (AO) method \cite{Alternating} by separating the optimization problem in \eqref{proposed_formulation_approximated} into two subproblems and address them iteratively.
The solution structure is shown in a flow chart in Fig. \ref{algorithm_flow_chart}.
In particular, subproblem 1 optimizes the user scheduling $\mathcal{S} = \{s_{k,k'}[n],\forall n,k,k'\}$
and the power allocation $\mathcal{P} = \{p_k[n], \forall n,k\}$ for a given
UAV's 3D trajectory $\mathcal{T} = \{\mathbf{t}[n], \forall n\}$, 3D flight velocity $\mathcal{V} = \{\mathbf{v}[n], \forall n\}$, and $\mathcal{Q} = \{\mathbf{q}_k[n], \forall n,k\}$;
Subproblem 2 optimizes the UAV's 3D trajectory $\mathcal{T} = \{\mathbf{t}[n], \forall n\}$, its 3D flight velocity $\mathcal{V} = \{\mathbf{v}[n], \forall n\}$, and $\mathcal{Q} = \{\mathbf{q}_k[n], \forall n,k\}$
for a given user scheduling $\mathcal{S} = \{s_{k,k'}[n],\forall n,k,k'\}$
and power allocation $\mathcal{P} = \{p_k[n], \forall n,k\}$.
Now, we study the solution of subproblem 1.

\begin{figure}[t]
\vspace*{-8mm}
  \centering
  \includegraphics[width=3.2 in]{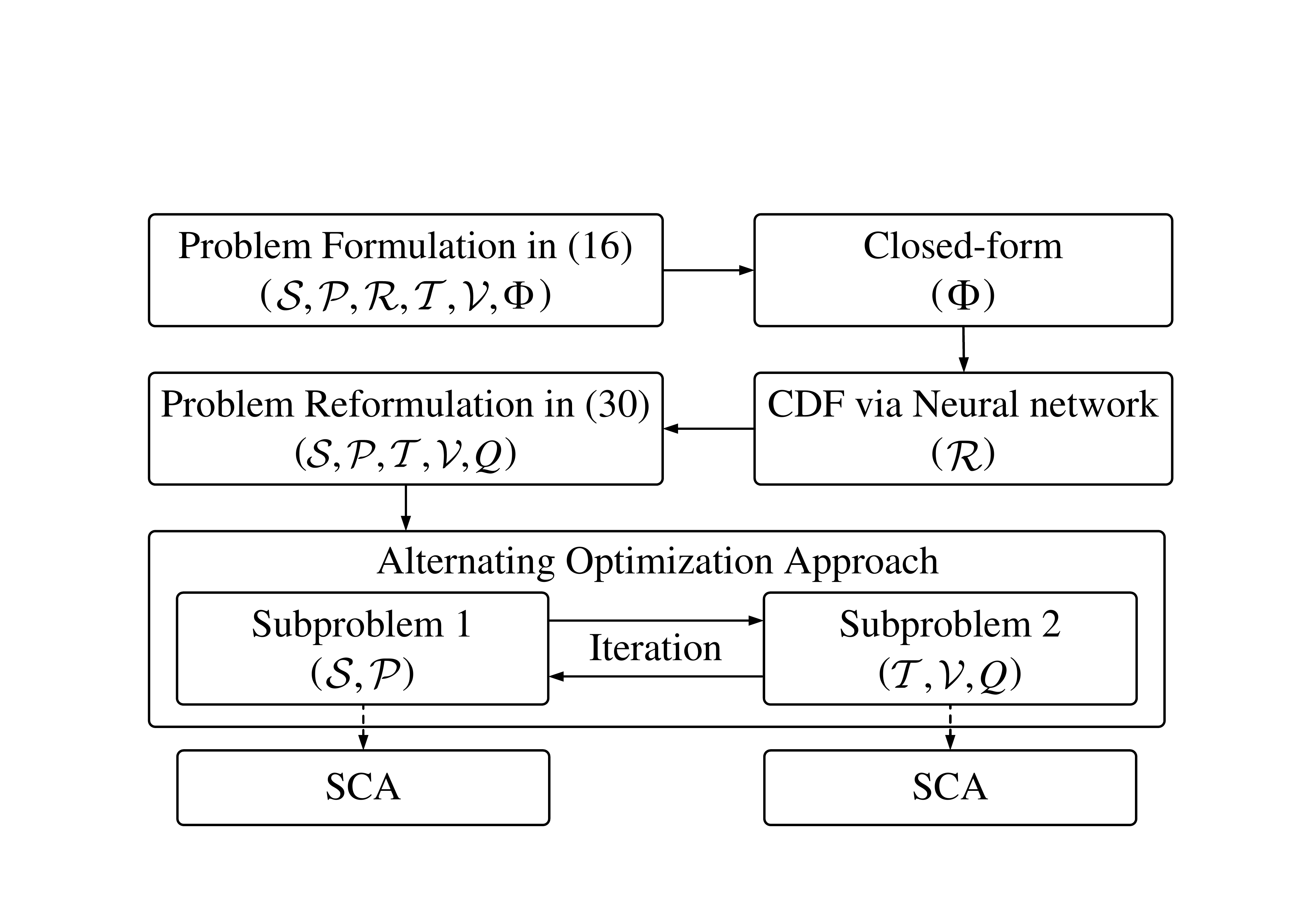}
  \vspace*{-12mm}
  \caption{A flow chart for the illustration of the proposed iterative algorithm.}
  \label{algorithm_flow_chart}
  \vspace*{-8mm}
\end{figure}

\vspace{-1mm}
\subsection{Subproblem 1: Optimizing User Scheduling and Power Allocation}
\vspace{-1mm}

In this subproblem, for any given UAV's trajectory and flight velocity, the user scheduling and power allocation can be formulated as:\vspace{-1mm}
\begin{align} \label{subproblem_1_original}
\underset{\mathcal{S},\mathcal{P}}{\mathrm{minimize}} \,\,& \frac{1}{N} \sum_{n=1}^N  P_{\mathrm{comm}}[n] + \frac{1}{N} \sum_{n=1}^N P_{\mathrm{fly}}[n] \\[-1mm]
\mathrm{s.t.} \,\,& \mathrm{C1-C4}, \overline{\mathrm{C5}}. \notag\\[-11mm]\notag
\end{align}

Note that constraint $\mathrm{C2}$ is an affine constraint w.r.t. the user scheduling $s_{k,k'}[n]$.
First, to address the nonconvexity of the problem, we handle the coupling between the paired user scheduling $s_{k,k'}[n]$ and transmit power allocation $p_k[n]$ variables by introducing one slack variable\footnote{Note that the slack variable with different subscripts have different physical meanings, i.e., $\tilde{p}_{k,k'\!,k}[n] = s_{k,k'}[n] p_k[n]$, $\tilde{p}_{k,k'\!,k'}[n] = s_{k,k'}[n] p_{k'}[n]$, $\tilde{p}_{k'\!,k,k}[n] = s_{k'\!,k}[n] p_{k}[n]$, $\tilde{p}_{k'\!,k,k'}[n] = s_{k'\!,k}[n] p_{k'}[n]$, and $\tilde{p}_{k,k,k}[n] = s_{k,k}[n] p_{k}[n]$ are the power allocation for GU $k$ and $k'$ when GU $k$ as the SIC user and GU $k'$ as the non-SIC user, the power allocation for GU $k$ and $k'$ when GU $k'$ as the SIC user and GU $k$ as the non-SIC user, and the power allocation for GU $k$ when it as the OMA user, respectively.
In this paper, we adopt $\tilde{p}_{k,k'\!,k}[n]$ to represent the slack variable to simplify the presentation.} $\tilde{p}_{k,k'\!,k}[n] = s_{k,k'}[n] p_k[n]$.
Then, by adopting the big-M formulation \cite{sun2018robust,lee2011mixed,9043712}, we introduce the following auxiliary constraints:\vspace{-1mm}
\begin{align}
\hspace{-2mm}\mathrm{C17}:\,& \tilde{p}_{k,k'\!,k}[n] \leq p_{k}[n], \forall n,k,k', \notag 
\hspace{1mm}\mathrm{C18}: \tilde{p}_{k,k'\!,k}[n] \leq s_{k,k'}[n] P_{\mathrm{peak}}, \forall n,k,k', \notag \\[-1mm]
\hspace{-2mm}\mathrm{C19}:\,& \tilde{p}_{k,k'\!,k}[n] \geq 0, \forall n,k,k',
\hspace{7.5mm}\mathrm{C20}: \tilde{p}_{k,k'\!,k}[n] \geq p_{k}[n] - (1-s_{k,k'}[n]) P_{\mathrm{peak}}, \forall n,k,k'. \hspace{-3mm} \\[-11mm]\notag
\end{align}
Then, we can rewrite the binary constraint $\mathrm{C1}$ in its equivalent form as 
\begin{equation}
\mathrm{C1a}: \sum_{n=1}^N \sum_{k=1}^K \sum_{k'=1}^K \left( s_{k,k'}[n] - (s_{k,k'}[n])^2 \right) \leq 0, \,\,
\mathrm{C1b}: 0 \leq s_{k,k'}[n] \leq 1, \forall n,k,k', 
\end{equation}
where variable $s_{k,k'}[n]$ is a continuous value between zero and one.
However, constraint $\mathrm{C1a}$ is a reverse convex function \cite{dinh2010local}.
To handle this non-convexity, we reformulate the problem formulation in \eqref{subproblem_1_original} based on \cite{Zhiqiang_DC,9043712} as its equivalent form: 
\begin{align}\label{subproblem_1_with_slack_variables}
& \underset{\mathcal{S},\mathcal{P},\tilde{\mathcal{P}}}{\mathrm{minimize}} \,\, \frac{1}{N} \sum_{n=1}^N \sum_{k=1}^K \bigg[ \eta \bigg( \sum_{\substack{k'=1\\ k'\neq k}}^K (\tilde{p}_{k,k'\!,k}[n]+ \tilde{p}_{k,k'\!,k'}[n]) + \tilde{p}_{k,k,k}[n] \bigg) \notag \\[-2mm]
& \hspace{17mm}+ \sum_{k'=1}^K \zeta\left( s_{k,k'}[n] - (s_{k,k'}[n])^2 \right) \bigg] + \frac{1}{N} \sum_{n=1}^N P_{\mathrm{fly}}[n] \\[-0.5mm]
& \mathrm{s.t.}\,\, \mathrm{C1b}, \mathrm{C2}, \mathrm{C17}-\mathrm{C20}, \notag \\[-0.55mm]
& \widetilde{\mathrm{C3}}: \tilde{p}_{k,k'\!,k}[n] \geq 0, \forall n,k,k'\!, \,\,
\widetilde{\mathrm{C4}}: \sum_{k=1}^K  \left( \sum_{k'\neq k}^K  (\tilde{p}_{k,k'\!,k} [n] + \tilde{p}_{k,k'\!,k'} [n]) + \tilde{p}_{k,k,k} [n] \right) \leq P_{\mathrm{peak}}, \forall n, \notag \\[-0.5mm]
& \widetilde{\mathrm{C5}}: \frac{1}{N} \sum_{n=1}^N \Bigg[ \sum_{\substack{k'=1\\
k'\neq k}}^K \tilde{R}_{k,k'}^{\mathrm{SIC}}[n] + \sum_{\substack{k'=1\\
k'\neq k}}^K  \left(\tilde{R}_{k'\!,k}^{\mathrm{I}}[n] - \tilde{R}_{k'\!,k}^{\mathrm{II}}[n]\right) + \tilde{R}_{k,k}^{\mathrm{OMA}}[n] \Bigg] \geq R_{\min_{k}}, \forall k, \notag  \\[-12mm]\notag
\end{align}
where \vspace{-0.5mm}
\begin{align}
\tilde{R}_{k,k'}^{\mathrm{SIC}}[n] &= s_{k,k'}[n] \log_2 \left( 1 + \frac{\tilde{p}_{k,k'\!,k}[n] f_k^{\mathrm{SIC}}[n]}{s_{k,k'}[n] \sigma_k^2} \right), \\[-0.5mm]
\tilde{R}_{k'\!,k}^{\mathrm{I}}[n] &= s_{k'\!,k}[n] \log_2 \left( \frac{(\tilde{p}_{k'\!,k,k}[n] + \tilde{p}_{k'\!,k,k'}[n]) f_{k}^{\mathrm{NSIC}}[n]}{s_{k'\!,k}[n]} + \sigma_k^2 \right),\\[-0.5mm]
\tilde{R}_{k'\!,k}^{\mathrm{II}}[n] &= s_{k'\!,k}[n] \log_2 \left( \frac{\tilde{p}_{k'\!,k,k'}[n] f_{k}^{\mathrm{NSIC}}[n]}{s_{k'\!,k}[n]} + \sigma_k^2 \right),\,\text{and} \\[-0.5mm]
\tilde{R}_{k,k}^{\mathrm{OMA}}[n] &= s_{k,k}[n] \log_2 \left( 1 + \frac{\tilde{p}_{k,k,k}[n] f_k^{\mathrm{OMA}}[n]}{s_{k,k}[n] \sigma_k^2} \right), \\[-11mm]\notag
\end{align}
$\tilde{\mathcal{P}} = \{\tilde{p}_{k,k'\!,k}[n], \forall n,k,k'\}$, and $\zeta\gg1$.
Note that the optimization problem in \eqref{subproblem_1_with_slack_variables} is still non-convex and the non-convexity arises from the objective function and constraint $\widetilde{\mathrm{C5}}$.
Thus, we handle the penalty terms in the objective function and $\tilde{R}_{k'\!,k}^{\mathrm{II}}[n]$ in nonconvex constraints $\widetilde{\mathrm{C5}}$ in problem \eqref{subproblem_1_with_slack_variables} via the iterative successive convex approximation (SCA) technique \cite{8876877,8891911}.
Specifically, for given $s^{j_1}_{k,k'}[n]$ and $p^{j_1}_{k,k'\!,k}[n]$ in the $j_1$-th iteration, an upper bound of the penalty term and $\tilde{R}_{k'\!,k}^{\mathrm{II}}[n]$ can be obtained by their first-order Taylor expansions as\vspace{-1mm}
\begin{align}
&\hspace{-2mm}s_{k,k'}[n] \!-\! (s_{k,k'}[n])^2 \leq (A_{k,k'}^{\mathrm{up}}[n])^{j_1} \!=\! s_{k,k'}[n] \!-\! (s_{k,k'}^{j_1}[n])^2 \!+\! 2 s_{k,k'}^{j_1}[n] (s_{k,k'}[n] \!-\! s_{k,k'}^{j_1}[n]) \,\,\text{and} \label{s_up} \hspace{-2mm}\\[-0.5mm]
&\hspace{-2mm}\tilde{R}_{k'\!,k}^{\mathrm{II}}[n]\leq (\tilde{R}_{k'\!,k}^{\mathrm{II,up}}[n])^{j_1} \notag \\[-0.5mm]
&\hspace{-2mm}= s_{k'\!,k}^{j_1}[n] \log_2 \left( \frac{\tilde{p}_{k'\!,k,k'}^{j_1}[n] f_{k}^{\mathrm{NSIC}}[n]}{s_{k'\!,k}^{j_1}[n]} \!+\! \sigma_k^2 \right) \!+ \log_2 \left( \frac{\tilde{p}_{k'\!,k,k'}^{j_1}[n] f_{k}^{\mathrm{NSIC}}[n]}{s_{k'\!,k}^{j_1}[n]} \!+\! \sigma_k^2 \right) \! (s_{k'\!,k}[n] \!-\! s_{k'\!,k}^{j_1}[n]) \notag \\[-0.2mm]
&\hspace{-2mm} - \frac{\tilde{p}_{k'\!,k,k'}^{j_1}[n] f_{k}^{\mathrm{NSIC}}[n] (s_{k'\!,k}[n] - s_{k'\!,k}^{j_1}[n])}{(\tilde{p}_{k'\!,k,k'}^{j_1}[n] f_{k}^{\mathrm{NSIC}}[n] + s_{k'\!,k}^{j_1}[n] \sigma_k^2) \ln 2} + \frac{s_{k'\!,k}^{j_1}[n] f_{k}^{\mathrm{NSIC}}[n] (\tilde{p}_{k'\!,k,k'}[n] - \tilde{p}_{k'\!,k,k'}^{j_1}[n])} {(\tilde{p}_{k'\!,k,k'}^{j_1}[n] f_{k}^{\mathrm{NSIC}}[n] + s_{k'\!,k}^{j_1}[n] \sigma_k^2) \ln 2}, \label{r_up}\\[-11mm]\notag
\end{align}
respectively.

After applying \eqref{s_up} and \eqref{r_up} to the transformed optimization problem in \eqref{subproblem_1_with_slack_variables}, we obtain a suboptimal solution by\vspace{-1mm}
\begin{align}\label{sub_1_final}
\underset{\mathcal{S},\mathcal{P},\tilde{\mathcal{P}}}{\mathrm{minimize}}  \, & \!\sum_{n=1}^N \!\sum_{k=1}^K \!\bigg[ \frac{1}{N} \!\bigg(\! \sum_{\substack{k'=1\\ k'\neq k}}^K (\tilde{p}_{k,k'\!,k}[n] \!+\! \tilde{p}_{k,k'\!,k'}[n]) \!+\! \tilde{p}_{k,k,k}[n] \!\bigg)  \!+\! \sum_{k'=1}^K \zeta ( A_{k,k'}^{\mathrm{up}}[n])^{j_1} \!\bigg] \!+\! \frac{1}{N} \!\sum_{n=1}^N \! P_{\mathrm{fly}}[n] \notag \\[-1mm]
\mathrm{s.t.} \,& \mathrm{C1b}, \mathrm{C2}, \widetilde{\mathrm{C3}}, \widetilde{\mathrm{C4}}, \mathrm{C17}-\mathrm{C20}, \\[-0.5mm]
\widetilde{\widetilde{\mathrm{C5}}}: \,& \frac{1}{N} \sum_{n=1}^N \Bigg[ \sum_{\substack{k'=1\\ k'\neq k}}^K \tilde{R}_{k,k'}^{\mathrm{SIC}}[n] + \sum_{\substack{k'=1\\ k'\neq k}}^K \left( \tilde{R}_{k'\!,k}^{\mathrm{I}}[n] - (\tilde{R}_{k'\!,k}^{\mathrm{II,up}}[n])^{j_1}\right) + \tilde{R}_{k,k}^{\mathrm{OMA}}[n] \Bigg] \geq R_{\min_{k}}, \forall k. \notag  \\[-11.5mm]\notag
\end{align}
Note that solving \eqref{sub_1_final} leads to an upper bound of the optimal objective value of problem \eqref{subproblem_1_with_slack_variables}.
Furthermore, in order to tighten the obtained upper bound, we iteratively update the feasible solution, $s_{k,k'}^{j_1}[n]$ and $\tilde{p}_{k'\!,k,k'}^{j_1}[n]$, by solving the optimization problem in \eqref{sub_1_final} with a standard convex optimization solver, such as CVX \cite{CVX} in the $j_1$-th iteration.
The proposed SCA-based algorithm is summarized in $\textbf{Algorithm \ref{alg_sub_1}}$ and the convergence of the algorithm to a suboptimal solution is guaranteed \cite{lower_bound_data_rate}.

\begin{table}[t] \vspace*{-6mm}
\scriptsize
\linespread{1}
\begin{algorithm} [H]
\caption{Proposed Algorithm for Handling Sub-problem 1} \label{alg_sub_1}
\begin{algorithmic} [1]

\STATE Initialize the convergence tolerance $\epsilon_1 \rightarrow 0$, the maximum number of iterations $I_{1,\max}$, the initial iteration index ${j_1} = 0$, the initial variables $\{s_{k,k'}^{j_1}[n],\tilde{p}_{k'\!,k,k'}^{j_1}[n]\}$, and the initial objective value $\tilde{P}_{\mathrm{total}}$

\REPEAT[Main Loop: SCA]

\STATE Set ${j_1} = {j_1} +1$ and $\{s_{k,k'}^{j_1}[n],\tilde{p}_{k'\!,k,k'}^{j_1}[n]\} = \{s_{k,k'}[n],\tilde{p}_{k'\!,k,k'}[n]\}$

\STATE Solving optimization problem in \eqref{sub_1_final} to obtain $\{s_{k,k'}[n],p_k[n],\tilde{p}_{k,k'\!,k}[n]\}$ and $\tilde{P}_{\mathrm{total}}$

\STATE Update $\tilde{P}_{\mathrm{total}}^{j_1}$ = $\tilde{P}_{\mathrm{total}}$

\UNTIL{${j_1} = I_{1,\max}$ or $\frac{|\tilde{P}_{\mathrm{total}}^{j_1} - \tilde{P}_{\mathrm{total}}^{({j_1}-1)}|}{\tilde{P}_{\mathrm{total}}^{j_1}} \leq \epsilon_1$}

\STATE Return $\{s_{k,k'}^*[n],p_{k}^*[n]\}$ = $\{s_{k,k'}[n],\tilde{p}_{k,k'\!,k}[n]\}$ and $\tilde{P}_{\mathrm{total}}^*$ = $\tilde{P}_{\mathrm{total}}^{j_1}$

\end{algorithmic}
\end{algorithm}
\vspace*{-16mm}
\end{table}

\vspace{-2mm}
\subsection{Subproblem 2: Optimizing UAV's 3D Trajectory and Flight Velocity}
\vspace{-1mm}

In this subproblem, for a given user scheduling and power allocation strategy, we can express the optimization problem as\vspace{-2.5mm}
\begin{align} \label{subproblem_2_original}
\underset{\mathcal{T},\mathcal{V},\mathcal{Q}}{\mathrm{minimize}} \,\, & \frac{1}{N} \sum_{n=1}^N P_{\mathrm{comm}}[n] + \frac{1}{N} \sum_{n=1}^N P_{\mathrm{fly}}[n] \\[-1mm]
\mathrm{s.t.}\,\, & \overline{\mathrm{C5}},\mathrm{C9}-\mathrm{C14}, \mathrm{C16}. \notag\\[-12mm]\notag
\end{align}
Note that the optimization problem in \eqref{subproblem_2_original} is nonconvex and the nonconvexity arises from constraint $\overline{\mathrm{C5}}$ and the function of the UAV's flight power consumption $P_{\mathrm{fly}}[n]$ w.r.t. $\mathbf{t}[n]$ and $\mathbf{v}[n]$, respectively.
Thus, to tackle these nonconvexities, we first introduce two slack variables $\nu[n]$ and $u_k^{i}[n]$ to rewrite the problem in \eqref{subproblem_2_original} into its equivalent form:\vspace{-1mm}
\begin{align} \label{subproblem_2_with_slack_variables}
\underset{\mathcal{T},\mathcal{V},\mathcal{Q},\Upsilon,\,\mathcal{U}}{\mathrm{minimize}} \,\, & \frac{1}{N} \sum_{n=1}^N P_{\mathrm{comm}}[n] + \frac{1}{N} \sum_{n=1}^N  \hat{P}_{\mathrm{fly}}[n] \\[-1mm]
\mathrm{s.t.} \,\,& \mathrm{C9}-\mathrm{C14},\mathrm{C16}, \notag \\[-1mm]
\widehat{\mathrm{C5}}: \,\,& \frac{1}{N} \sum_{n=1}^N \Bigg( \sum_{\substack{k'=1\\ k'\neq k}}^K \hat{R}_{k,k'}^{\mathrm{SIC}}[n] + \sum_{\substack{k'=1\\ k'\neq k}}^K \left(\hat{R}_{k'\!,k}^{\mathrm{I}}[n] - \hat{R}_{k'\!,k}^{\mathrm{II}}[n]\right) + \hat{R}_{k,k}^{\mathrm{OMA}}[n] \Bigg) \geq R_{\min_k}, \!\forall k, \notag \\[-1mm]
\mathrm{C21}: \,\,& v_x^2[n] + v_y^2[n] \geq \nu^2[n], \forall n,
\hspace{20.6mm} \mathrm{C22}: \nu[n] \geq 0, \forall n, \notag \\[-0.5mm]
\mathrm{C23}: \,\,& u_k^{i}[n] \geq (\mathbf{w}_{3}^{i})^{\mathrm H} \left[ \mathbf{w}_{2}^{i} \left[\mathbf{w}_{1}^{i} \mathbf{q}_k[n] + \mathbf{b}_{1}^{i} \right]^+ + \mathbf{b}_{2}^{i} \right]^+ + b_3^i, \forall n,k,i, \notag \\[-11mm]\notag
\end{align}
where $\Upsilon = \{\nu[n], \forall n\}$, $\mathcal{U} = \{u^{i}_k[n], \forall k,n\}$,
\begin{align}
\hspace{-3mm} \hat{R}_{k,k'}^{\mathrm{SIC}}[n] &= s_{k,k'}[n] \log_2 \!\left(\! 1 \!+\! \frac{\tilde{p}_{k,k'\!,k}[n]}{u_k^{\mathrm{SIC}}[n] \sigma_k^2} \right)\!, \label{data_rate_bar}
\hat{R}_{k'\!,k}^{\mathrm{I}}[n] = s_{k'\!,k}[n] \log_2 \!\left(\! \frac{\tilde{p}_{k'\!,k,k'\!}[n] \!+\! \tilde{p}_{k'\!,k,k}[n]}{u_k^{\mathrm{NSIC}}[n]} \!+\! \sigma_k^2 \!\right)\!\!, \hspace{-3mm} \\
\hspace{-3mm} \hat{R}_{k'\!,k}^{\mathrm{II}}[n] &= s_{k'\!,k}[n] \log_2 \left(\frac{\tilde{p}_{k'\!,k,k'\!}[n]}{u_k^{\mathrm{NSIC}}[n]} \!+\! \sigma_k^2\right)\!,
\, \hat{R}_{k,k}^{\mathrm{OMA}}[n]= s_{k,k}[n] \log_2 \left( 1 \!+\! \frac{\tilde{p}_{k,k,k}[n]}{u_k^{\mathrm{OMA}}[n] \sigma_k^2} \right)\!, \,  \text{and} \hspace{-3mm} \\
\hspace{-3mm} \hat{P}_{\mathrm{fly}}[n] &= P_o \left( 1 + \frac{3 (v_x^2[n] + v_y^2[n]) }{\Omega^2 r^2} \right) \!+\! \frac{P_i v_0}{\nu[n]} \!+\! \frac{1}{2} d_0 \rho s A_{\mathrm r} (v_x^2[n] + v_y^2[n])^{3/2} + G v_{\mathrm{z}}[n].\\[-11mm]\notag
\end{align}
Note that the additional inequality constraints $\mathrm{C21}-\mathrm{C25}$ in problem \eqref{subproblem_2_with_slack_variables} are all active at the optimal point.
Thus, the formulated problems in \eqref{subproblem_2_original} and \eqref{subproblem_2_with_slack_variables} are equivalent to each other.
Moreover, the nonconvexity of constraint $\mathrm{C23}$ in problem \eqref{subproblem_2_with_slack_variables} is attributed to the vector parameter $\mathbf{w}_2^i$ involving both positive and negative values, although the ReLU function is convex w.r.t. $\mathbf{q}_k[n]$.
To address this issue, we introduce two indicator variables $\mathbf{a}_{1,k}^i[n] \in \mathbb{R}^{200\times200}$ and $\mathbf{a}_{2,k}^i[n] \in \mathbb{R}^{100\times100}$ as\vspace{-1mm}
\begin{equation}
\mathrm{C24}: \mathbf{a}_{1,k}^i[n] \in \{0,1\}, \forall n,k,i, \,\,
\mathrm{C25}: \mathbf{a}_{2,k}^i[n] \in \{0,1\}, \forall n,k,i,\\[-1mm]
\end{equation}
where $\mathbf{a}_{1,k}^i[n] = 1$ when $\mathbf{w}_{1}^{i} \mathbf{q}_k[n] + \mathbf{b}_{1}^{i}> 0$. Otherwise, $\mathbf{a}_{1,k}^i[n] = 0$.
Similarly, $\mathbf{a}_{2,k}^i[n] = 1$ when $\mathbf{w}_{2}^{i} \left( \mathbf{w}_{1}^{i} \mathbf{q}_k[n] + \mathbf{b}_{1}^{i} \right) + \mathbf{b}_{2}^{i} > 0$. Otherwise, $\mathbf{a}_{2,k}^i[n] = 0$.
Thus, constraint $\mathrm{C23}$ can be rewritten as\vspace{-1mm}
\begin{equation}
\widehat{\mathrm{C23}}: u_k^i[n] \geq (\mathbf{w}_{3}^{i})^{\mathrm H} \mathbf{a}_{2,k}^i[n] \left( \mathbf{w}_{2}^{i} \mathbf{a}_{1,k}^i[n] \left(\mathbf{w}_{1}^{i} \mathbf{q}_k[n] + \mathbf{b}_{1}^{i} \right) + \mathbf{b}_{2}^{i} \right) + b_{3}^{i}, \forall n,k,i. \\[-1mm]
\end{equation}
Note that similar to the solution of subproblem 1, we handle the coupling of $\mathbf{a}_{1,k}^i[n]$, $\mathbf{a}_{2,k}^i[n]$, and $\mathbf{q}_k[n]$ by introducing two slack variables $ \hat{\mathbf{q}}_k^i[n] = \mathbf{a}_{1,k}^i[n] \left(\mathbf{w}_{1}^{i} \mathbf{q}_k[n] + \mathbf{b}_{1}^{i} \right) \in \mathbb{R}^{200\times1}$ and $ \hat{\hat{\mathbf{q}}}_k^i[n] = \mathbf{a}_{2,k}^i[n] \left(\mathbf{w}_{2}^{i} \hat{\mathbf{q}}_k^i[n] + \mathbf{b}_{2}^{i} \right) \in \mathbb{R}^{100\times1}$, and introduce the following constraints based on big-M formulation:\vspace{-1mm}
\begin{align}
& \mathrm{C26}: \hat{\mathbf{q}}_k^i[n] \geq \mathbf{w}_{1}^{i} \mathbf{q}_k[n] + \mathbf{b}_{1}^{i}, \forall n,k,i, \hspace{10mm}
\mathrm{C27}: \hat{\mathbf{q}}_k^i[n] \geq 0, \!\forall n,k,i, \notag \\
& \mathrm{C28}: \hat{\mathbf{q}}_k^i[n] \leq \mathbf{a}_{1,k}^i[n] \left(\mathbf{w}_{1}^{i} \left[ \mathbf{t}_{\max}; \mathbf{l}_k \right] + \mathbf{b}_{1}^{i} \right), \forall n,k,i, \notag \\
& \mathrm{C29}: \hat{\mathbf{q}}_k^i[n] \leq \mathbf{w}_{1}^{i} \mathbf{q}_k[n] + \mathbf{b}_{1}^{i} + (1 - \mathbf{a}_{1,k}^i[n]) \left(\mathbf{w}_{1}^{i} \left[ \mathbf{t}_{\max}; \mathbf{l}_k \right] + \mathbf{b}_{1}^{i} \right), \forall n,k,i, \notag \\
& \mathrm{C30}: \hat{\hat{\mathbf{q}}}_k^i[n] \geq \mathbf{w}_{2}^{i} \hat{\mathbf{q}}_k^i[n] + \mathbf{b}_{2}^{i}, \forall n,k,i, \hspace{10mm}
\mathrm{C31}: \hat{\hat{\mathbf{q}}}_k^i[n] \geq 0, \!\forall n,k,i, \notag \\
& \mathrm{C32}: \hat{\hat{\mathbf{q}}}_k^i[n] \leq \mathbf{a}_{2,k}^i[n] \left( \mathbf{w}_{2}^{i} \left(\mathbf{w}_{1}^{i} \left[ \mathbf{t}_{\max}; \mathbf{l}_k \right] + \mathbf{b}_{1}^{i} \right) + \mathbf{b}_{2}^{i} \right), \forall n,k,i, \notag \\
& \mathrm{C33}: \hat{\hat{\mathbf{q}}}_k^i[n] \leq \mathbf{w}_{2}^{i} \hat{\mathbf{q}}_k^i[n] + \mathbf{b}_{2}^{i} + (1 - \mathbf{a}_{2,k}^i[n]) \left( \mathbf{w}_{2}^{i} \left(\mathbf{w}_{1}^{i} \left[ \mathbf{t}_{\max}; \mathbf{l}_k \right] + \mathbf{b}_{1}^{i} \right) + \mathbf{b}_{2}^{i} \right), \forall n,k,i. \\[-11mm]\notag
\end{align}
Then, we rewrite the binary constraints $\mathrm{C24}$ and $\mathrm{C25}$ in their equivalent forms as\vspace{-1mm}
\begin{align}
\mathrm{C24a}: \mathbf{a}_{1,k}^i[n] - \left(\mathbf{a}_{1,k}^i[n]\right)^2 \leq 0, \forall n,k,i, \hspace{4mm}
\mathrm{C24b}: 0 \leq \mathbf{a}_{1,k}^i[n] \leq 1, \forall n,k,i, \notag\\
\mathrm{C25a}: \mathbf{a}_{2,k}^i[n] - \left(\mathbf{a}_{2,k}^i[n]\right)^2 \leq 0, \forall n,k,i, \hspace{4mm}
\mathrm{C25b}: 0 \leq \mathbf{a}_{2,k}^i[n] \leq 1, \forall n,k,i, \\[-11mm]\notag
\end{align}
respectively.
Next, we handle the nonconvex constraints $\widehat{\mathrm{C5}}$, $\mathrm{C21}$, $\mathrm{C24a}$, and $\mathrm{C25a}$ via SCA.
In particular, for a given feasible solution, $(u_k^{i}[n])^{j_2}$, $v_x^{j_2}[n]$, $v_y^{j_2}[n]$, $(\mathbf{a}_{1,k}^i[n])^{j_2}$, and $(\mathbf{a}_{2,k}^i[n])^{j_2}$ in the $j_2$-th iteration, the lower bound function of $\widehat{\mathrm{C5}}$, $\mathrm{C21}$, $\mathrm{C24a}$, and $\mathrm{C25a}$ can be constructed based on their first-order Taylor expansions \cite{lower_bound_data_rate}, respectively, which are given by\vspace{-1mm}
\begin{align}
&\hspace{-2mm} \hat{R}_{k,k'}^{\mathrm{SIC}}[n] \geq (\hat{R}_{k,k'}^{\mathrm{SIC,lb}}[n])^{j_2} \label{data_rate_lower_bound} \\
&\hspace{-2mm} = s_{k,k'}[n] \log_2 \left( 1 + \frac{\tilde{p}_{k,k'\!,k}[n]}{(u_k^{\mathrm{SIC}}[n])^{j_2} \sigma_k^2} \right) - \frac{s_{k,k'}[n] \tilde{p}_{k,k'\!,k}[n] (u_k^{\mathrm{SIC}}[n] - (u_k^{\mathrm{SIC}}[n])^{j_2})}{(u_k^{\mathrm{SIC}}[n])^{j_2} ((u_k^{\mathrm{SIC}}[n])^{j_2} \sigma_k^2 + \tilde{p}_{k,k'\!,k}[n]) \ln 2}, \notag \hspace{-2mm} \\
&\hspace{-2mm} \hat{R}_{k'\!,k}^{\mathrm I}[n] \geq (\hat{R}_{k'\!,k}^{\mathrm{I,lb}}[n])^{j_2} = s_{k'\!,k}[n] \log_2 \left( \frac{\tilde{p}_{k'\!,k,k}[n] + \tilde{p}_{k'\!,k,k'\!}[n]}{(u_k^{\mathrm{NSIC}}[n])^{j_2}} + \sigma_k^2 \right) \\
&\hspace{-2mm} - \frac{s_{k'\!,k}[n] (\tilde{p}_{k'\!,k,k}[n] + \tilde{p}_{k'\!,k,k'\!}[n]) (u_k^{\mathrm{NSIC}}[n] - (u_k^{\mathrm{NSIC}}[n])^{j_2})}{(u_k^{\mathrm{NSIC}}[n])^{j_2} ((u_k^{\mathrm{NSIC}}[n])^{j_2} \sigma_k^2 + \tilde{p}_{k'\!,k,k}[n] + \tilde{p}_{k'\!,k,k'\!}[n]) \ln 2}, \notag \\
&\hspace{-2mm} \hat{R}_{k,k}^{\mathrm{OMA}}[n] \geq (\hat{R}_{k,k}^{\mathrm{OMA,lb}}[n])^{j_2} \notag \\
&\hspace{-2mm} = s_{k,k}[n] \log_2 \left( 1 + \frac{\tilde{p}_{k,k,k}[n]}{(u_k^{\mathrm{OMA}}[n])^{j_2} \sigma_k^2} \right) - \frac{s_{k,k}[n] \tilde{p}_{k,k,k}[n] (u_k^{\mathrm{OMA}}[n] - (u_k^{\mathrm{OMA}}[n])^{j_2})}{(u_k^{\mathrm{OMA}}[n])^{j_2} ((u_k^{\mathrm{OMA}}[n])^{j_2} \sigma_k^2 \!+\! \tilde{p}_{k,k,k}[n]) \ln 2}, \notag\\
&\hspace{-2mm} v_x^2[n] + v_y^2[n] \geq (v_x^{j_2}[n])^2 + (v_y^{j_2}[n])^2 + 2 v_x^{j_2}[n] (v_x[n] - v_x^{j_2}[n]) + 2 v_y^{j_2}[n] (v_y[n] - v_y^{j_2}[n]), \hspace{-3mm} \label{velocity_lower_bound}\\[-12mm] \notag
\end{align}
\begin{align}
&\hspace{-2mm} \mathbf{a}_{1,k}^i[n] - \left(\mathbf{a}_{1,k}^i[n]\right)^2 \leq \mathbf{a}_{1,k}^i[n] - \left((\mathbf{a}_{1,k}^i[n])^{j_2}\right)^2 + 2 (\mathbf{a}_{1,k}^i[n])^{j_2} \left( \mathbf{a}_{1,k}^i[n] - (\mathbf{a}_{1,k}^i[n])^{j_2}\right), \, \text{and}\hspace{-2mm} \\
&\hspace{-2mm} \mathbf{a}_{2,k}^i[n] - \left(\mathbf{a}_{2,k}^i[n]\right)^2 \leq \mathbf{a}_{2,k}^i[n] - \left((\mathbf{a}_{2,k}^i[n])^{j_2}\right)^2 + 2 (\mathbf{a}_{2,k}^i[n])^{j_2} \left( \mathbf{a}_{2,k}^i[n] - (\mathbf{a}_{2,k}^i[n])^{j_2}\right), \label{indicator_upper_bound} \\[-11mm]\notag
\end{align}
respectively.

\begin{table}[t] \vspace*{-6mm}
\scriptsize
\linespread{1}
\begin{algorithm} [H]
\caption{Proposed Algorithm for Handling Sub-problem 2} \label{alg_sub_2}
\begin{algorithmic} [1]

\STATE Initialize the convergence tolerance $\epsilon_2 \rightarrow 0$, the maximum number of iterations $I_{2,\max}$, the initial iteration index ${j_2} = 0$, the initial variables $\{v_{\mathrm x}^{j_2}[n],v_{\mathrm y}^{j_2}[n]\}$, and the initial objective value $\hat{P}_{\mathrm{total}}$

\REPEAT[Main Loop: SCA]

\STATE Set ${j_2} = {j_2} + 1$, $\{(u_k^{i}[n])^{j_2}, v_{\mathrm x}^{j_2}[n],v_{\mathrm y}^{j_2}[n]\} \!=\! \{u_k^{i}[n], v_{\mathrm x}[n],v_{\mathrm y}[n]\}$

\STATE Solving optimization problem in \eqref{subproblem_2_final} to obtain $\{\mathbf{t}[n], \mathbf{v}[n], \nu[n], u_k^{i}[n]\}$ and $\hat{P}_{\mathrm{total}}$

\STATE Update $\hat{P}_{\mathrm{total}}^{j_2} = \hat{P}_{\mathrm{total}}$,

\UNTIL{${j_2} = I_{2,\max}$ or $\frac{|\hat{P}_{\mathrm{total}}^{j_2} - \hat{P}_{\mathrm{total}}^{({j_2}-1)}|}{\hat{P}_{\mathrm{total}}^{j_2}} \leq \epsilon_2$}

\STATE Return $\{\mathbf{t}^*[n], \mathbf{v}^*[n]\}$ = $\{\mathbf{t}[n], \mathbf{v}[n]\}$ and $\hat{P}_{\mathrm{total}}^*$ = $\hat{P}_{\mathrm{total}}^{j_2}$

\end{algorithmic}
\end{algorithm}
\vspace*{-11mm}
\end{table}

\begin{table}[t] \vspace*{-6mm}
\scriptsize
\linespread{1}
\begin{algorithm} [H]
\caption{Overall Algorithm for Addressing Problem \eqref{proposed_formulation_origion}} \label{alg_overall}
\begin{algorithmic} [1]

\STATE Initialize the convergence tolerance $\epsilon_3 \rightarrow 0$, the maximum number of iterations $I_{3,\max}$, the initial iteration index $j_3 = 0$, and the initial trajectory $\{\mathbf{t}[n], \mathbf{v}[n]\}$

\REPEAT

\STATE Set $j_3 = j_3+1$

\STATE Using $\textbf{Algorithm \ref{alg_sub_1}}$ to obtain a suboptimal result $\tilde{P}_{\mathrm{total}}$, $\{s_{k}[n], p_k[n]\}$, given the UAV's trajectory and flight velocity $\{\mathbf{t}[n], \mathbf{v}[n]\}$

\STATE Using $\textbf{Algorithm \ref{alg_sub_2}}$ to obtain a suboptimal result $\hat{P}_{\mathrm{total}}$, $\{\mathbf{t}[n], \mathbf{v}[n]\}$, given the resource allocation $\{s_{k}[n], p_k[n]\}$

\STATE Update $\hat{P}_{\mathrm{total}}^{j_3} = \hat{P}_{\mathrm{total}}$

\UNTIL{$j_3=I_{3,\max}$ or $\frac{|\hat{P}_{\mathrm{total}}^{j_3} - \hat{P}_{\mathrm{total}}^{(j_3-1)}|}{\hat{P}_{\mathrm{total}}^{j_3}} \leq \epsilon_2$}

\RETURN $\{s_{k}^*[n], p_k^*[n], \mathbf{t}^*[n], \mathbf{v}^*[n]\}$ = $\{s_{k}[n], p_k[n], \mathbf{t}[n], \mathbf{v}[n]\}$ and $P_{\mathrm{total}}^*=\hat{P}_{\mathrm{total}}^{j_3}$

\end{algorithmic}
\end{algorithm}
\vspace*{-15mm}
\end{table}

Now, applying the lower bounds in \eqref{data_rate_lower_bound}--\eqref{indicator_upper_bound} to \eqref{subproblem_2_with_slack_variables} yields the following convex optimization problem:\vspace{-0.5mm}
\begin{align} \label{subproblem_2_final}
&\underset{\mathcal{T},\mathcal{V},\mathcal{Q},\Upsilon,\,\mathcal{U}, \mathcal{A},\hat{\mathcal{Q}}} {\mathrm{minimize}} \ \frac{1}{N} \sum_{n=1}^N P_{\mathrm{comm}}[n] + \frac{1}{N} \sum_{n=1}^N \hat{P}_{\mathrm{fly}}[n] \\[-0.5mm]
& \mathrm{s.t.} \,\mathrm{C9}-\mathrm{C14}, \mathrm{C16},\mathrm{C22}, \mathrm{C24b},\mathrm{C25b},\mathrm{C26}-\mathrm{C33}, \notag \\[-0.5mm]
&\widehat{\widehat{\mathrm{C5}}}: \frac{1}{N} \sum_{n=1}^N \!\Bigg( \sum_{\substack{k'=1\\ k'\neq k}}^K  (\hat{R}_{k,k'}^{\mathrm{SIC,lb}}[n])^{j_2} \!+\! \sum_{\substack{k'=1\\ k'\neq k}}^K \! \left( (\hat{R}_{k'\!,k}^{\mathrm{I,lb}}[n])^{j_2} \!- \hat{R}_{k'\!,k}^{\mathrm{II}}[n] \right) \!+\! (\hat{R}_{k,k}^{\mathrm{OMA,lb}}[n])^{j_2} \!\Bigg) \!\geq\! R_{\min_k}, \forall k, \notag \\[-1mm]
&\widehat{\mathrm{C21}}: (v_x^{j_2}[n])^2 \!+\! (v_y^{j_2}[n])^2 \!+\! 2 v_x^{j_2}[n] (v_x[n]-v_x^{j_2}[n]) \!+\! 2 v_y^{j_2}[n] (v_y[n] - v_y^{j_2}[n]) \!\geq\! \nu^2[n], \forall n, \notag \\[-0.5mm]
&\widehat{\widehat{\mathrm{C23}}}: u_k^i[n] \geq (\mathbf{w}_{3}^{i})^{\mathrm H} \hat{\hat{\mathbf{q}}}_k^i[n] + b_{3}^{i}, \forall n,k,i, \notag \\[-0.5mm]
&\widehat{\mathrm{C24a}}: \mathbf{a}_{1,k}^i[n] - \left((\mathbf{a}_{1,k}^i[n])^{j_2}\right)^2 + 2 (\mathbf{a}_{1,k}^i[n])^{j_2} \left( \mathbf{a}_{1,k}^i[n] - (\mathbf{a}_{1,k}^i[n])^{j_2}\right) \leq 0, \forall n,k,i, \notag \\[-0.5mm]
&\widehat{\mathrm{C25a}}: \mathbf{a}_{2,k}^i[n] - \left((\mathbf{a}_{2,k}^i[n])^{j_2}\right)^2 + 2 (\mathbf{a}_{2,k}^i[n])^{j_2} \left( \mathbf{a}_{2,k}^i[n] - (\mathbf{a}_{2,k}^i[n])^{j_2}\right) \leq 0, \forall n,k,i, \notag\\[-11mm]\notag
\end{align}
where $\mathcal{A} = \{\mathbf{a}_{1,k}^i[n], \mathbf{a}_{2,k}^i[n], \forall n,k,i\}$ and $\hat{\mathcal{Q}} = \{\hat{\mathbf{q}}_k^i[n], \hat{\hat{\mathbf{q}}}_k^i[n], \forall n,k,i\}$.
Note that similar to the solution of subproblem 1, the optimization problem in \eqref{subproblem_2_final} is convex formulations, which can be easily solved by CVX \cite{CVX}.
The proposed algorithm is summarized in $\textbf{Algorithm \ref{alg_sub_2}}$.

\vspace{-1mm}
\subsection{Overall Algorithm}
\vspace{-1mm}

The overall algorithm for solving the two subproblems in \eqref{subproblem_1_original} and \eqref{subproblem_2_original} iteratively are summarized in $\textbf{Algorithm \ref{alg_overall}}$.
The convergence of the overall proposed algorithm to a stationary point monotonically can be guaranteed due to the compactness of the feasible solution set in \eqref{proposed_formulation_origion} and the nonincreasing objective value over iterations.
Besides, we adopt the solution of subproblem 2 as an input for subproblem 1 over iterations while solving the subproblem in \eqref{subproblem_1_original} and \eqref{subproblem_2_original} iteratively.
{Besides, the overall iterative algorithm can be shown to converge to a suboptimal solution of the optimization problem in \eqref{proposed_formulation_origion}, c.f.\cite{Alternating,wu2018joint,opial1967weak,5519540}}.

Furthermore, the computational complexity of the proposed suboptimal algorithm is given by \cite{nesterov1994interior,nonlinear}\vspace{-1mm}
\begin{equation}
\mathcal{O} \bigg( I_{1,\max} \bigg( \underset{\text{Subproblem 1}}{\underbrace{\mathcal{M}_1 \mathcal{N}_1^2 \times \sqrt{\mathcal{M}_1} \log \bigg(\frac{1}{\Delta_1}\bigg) }} + \underset{\text{Subproblem 2}} {\underbrace{\mathcal{M}_2 \mathcal{N}_2^2 \times I_{2,\max} \sqrt{\mathcal{M}_2} \log \bigg(\frac{1}{\Delta_2} \bigg) }} \bigg) \bigg),\\[-1mm]
\end{equation}
where $\mathcal{M}_1 = 6NK^2+2N+K$, $\mathcal{N}_1 = NK^3+NK^2+NK$, $\mathcal{M}_2 = 40NK+6N+K+1$, and $\mathcal{N}_2 = 24NK+(100^2+200^2)3 NK+7N$ represent the number of inequalities and the number of variables of subproblem 1 and subproblem 2, respectively.
Besides, $\Delta_1>0$ and $\Delta_2>0$ denote the thresholds of convergence tolerance of subproblem 1 and subproblem 2, respectively.
Note that we did not take into account the computational complexity of the adopted DNN approach to approximate the outage-guaranteed channel gain when calculating the complexity of the algorithm, as it is computed for once before the execution of the algorithm when the system parameters are determined.
Thus, the computational complexity of the proposed suboptimal algorithm is with polynomial time which is suitable for fast implementation \cite{weiss2000fast}.

\begin{table}[t] \vspace{-2mm}
\scriptsize
\linespread{1.14}
\caption{Simulation parameters \cite{EE_fixed_wing,8811733}.} \label{simulation_setting}
\begin{center}
\vspace{-8mm}
\begin{tabular}{ c | c | c | c | c | c | c | c}
  \hline			
  Notations                 & Simulation value          & Notations                     & Simulation value     	
  & Notations               & Simulation value          & Notations                     & Simulation value \\ \hline
  $K$              		    & 1 $\sim$ 10               & $\mathbf{t}_0$     		    & $[0;0;150]$ m			
  & $V_{\max}$    		    & 30 m/s                    & $\varepsilon_k^{\mathrm{SIC}}$				    & 0.01 \\
  $N$                       & 500 $\sim$ 1,000              & $\mathbf{t}_{\mathrm{F}}$     & $[500;500;150]$ m 	
  & $V_{\mathrm{acc}}$ 	    & 4 m/$\text{s}^{\text 2}$  & $\varepsilon_k^{\mathrm{NSIC}}$				    & 0.01 \\
  $\sigma_k^2$                & -160 dBm/Hz               & $\mathbf{t}_{\min}$    	    & $[0;0;100]$ m			
  & $\tau$                  & 0.1 s                       & $\varepsilon_k^{\mathrm{OMA}}$				    & 0.01 \\
  $\beta_0$                 & -50 dBW    		        & $\mathbf{t}_{\max}$    	    & $[500;500;300]$ m 	
  & $P_{\mathrm{peak}}$     & 36 dBm                    & $\alpha^{\mathrm{AR}}$        & 2 \\
  $\lambda_{\mathrm{C}}$	& 0.1 m                    & $\mathbf{l}_1$    		    & $[300;150;0]$ m 		
  & $A_1$					& 0 dB	                    & $\alpha^{\mathrm{RG}}$        & 3.6 \\
  $M_{\mathrm R}$           & 100              & $\mathbf{l}_2$    		    & $[50;400;0]$ m		
  & $A_2$					& 6.43 dB                   & $\alpha^{\mathrm{AG}}$        & 3.6 \\
  $\mathbf{l}_{\mathrm R}$  & $[0;400;30]$ m		    & $\mathbf{l}_3$  			    & $[100;450;0]$ m 		
  & $R_{\min_k}$              & 0.5 $\sim$ 5 bits/s/Hz    & $\kappa^{\mathrm{RG}}$        & 2 dB  \\
  $I_{1,\max}$  			    & 10 		
  & $I_{2,\max}$            & 10    &  $\kappa_{\min}$       	        & 0 dB    	        &  $\kappa_{\max}$       	        & 30 dB \\
  \hline
\end{tabular}
\end{center}
\vspace{-13mm}
\end{table}

\vspace{-2mm}
\section{Numerical Results}
\vspace{-1mm}

In this section, we discuss the system performance of the proposed scheme (PS) based on the following simulation results.
The simulation parameters are summarized in Table \ref{simulation_setting}.
Generally, we set $K=3$ for illustration to unveil the assistance brought by the IRS to the UAV communications.
{In addition, the initial trajectory of the UAV for $\textbf{Algorithm \ref{alg_overall}}$ is set as a piecewise linear flight locus at a fixed altitude of $100$ meters which the UAV passes by all the GUs in between the starting point and the final point with a constant velocity.}
In order to illustrate the performance gain of the IRS to the UAV communications, we compare the system performance of the PS with different numbers of the IRS elements and some baseline schemes.
In particular, we compare the PS with five baseline schemes:
(a) \emph{OMA consideration only (OMA)}, where the UAV only serve one GU at each time slot for the OMA scheme and all the other setups remain the same as the PS;
(b) \emph{No IRS consideration (NI)}, which removes the IRS from the considered UAV communication system;
(c) \emph{Constant flight altitude of the UAV (CFA)}, where the UAV operates at a constant altitude (i.e., $100$ meters) and only the horizontal trajectory of the UAV is optimized;
(d) \emph{Perfect CSI (PCSI)}, where the signal model is based on perfect known CSI and all channels are pure LoS dominated;
(e) \emph{Straight trajectory of the UAV (ST)}, where the UAV flies with a straight line trajectory from the initial location to the final location with a constant flight velocity, i.e., $11$ m/s.
Note that the corresponding resource allocation for NI, CFA, and ST is a subcase of the PS which can be obtained by $\textbf{Algorithm \ref{alg_overall}}$ with some straightforward modifications.

\vspace{-2mm}
\subsection{Convergence of the Proposed Scheme and Baseline Schemes}
\vspace{-1mm}

\begin{figure}[t]\vspace*{-2mm}
  \centering
  \begin{minipage}[t]{0.49\textwidth}
      \centering
      \hspace*{-5.3mm} \includegraphics[width=3.3 in]{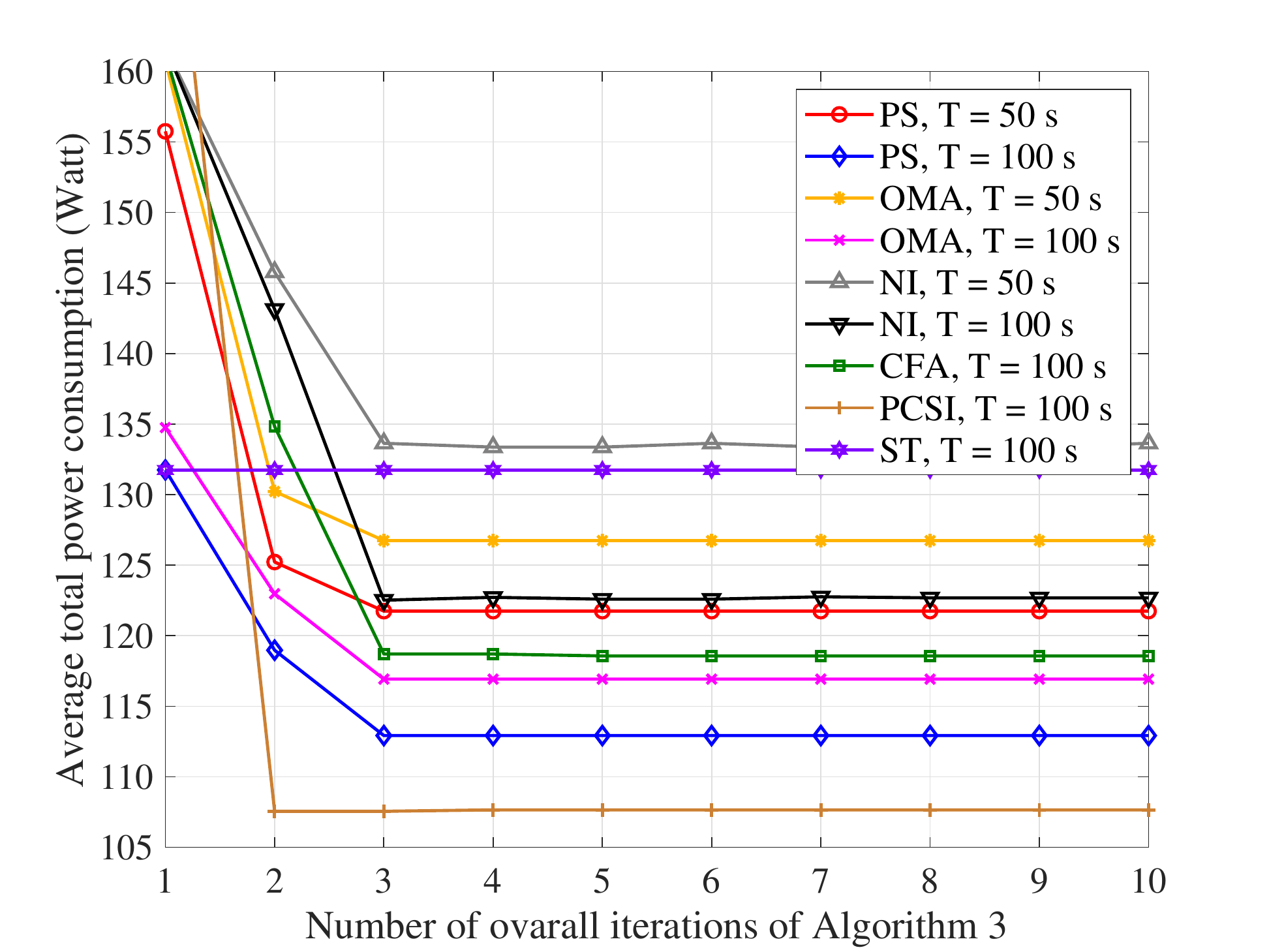}
      \vspace*{-9mm}
      \caption{Convergence of the PS and baseline schemes for different setups.}
      \label{convergence}
  \end{minipage}\,\,
  \begin{minipage}[t]{0.49\textwidth}
      \centering
      \hspace*{-6mm} \includegraphics[width=3.3 in]{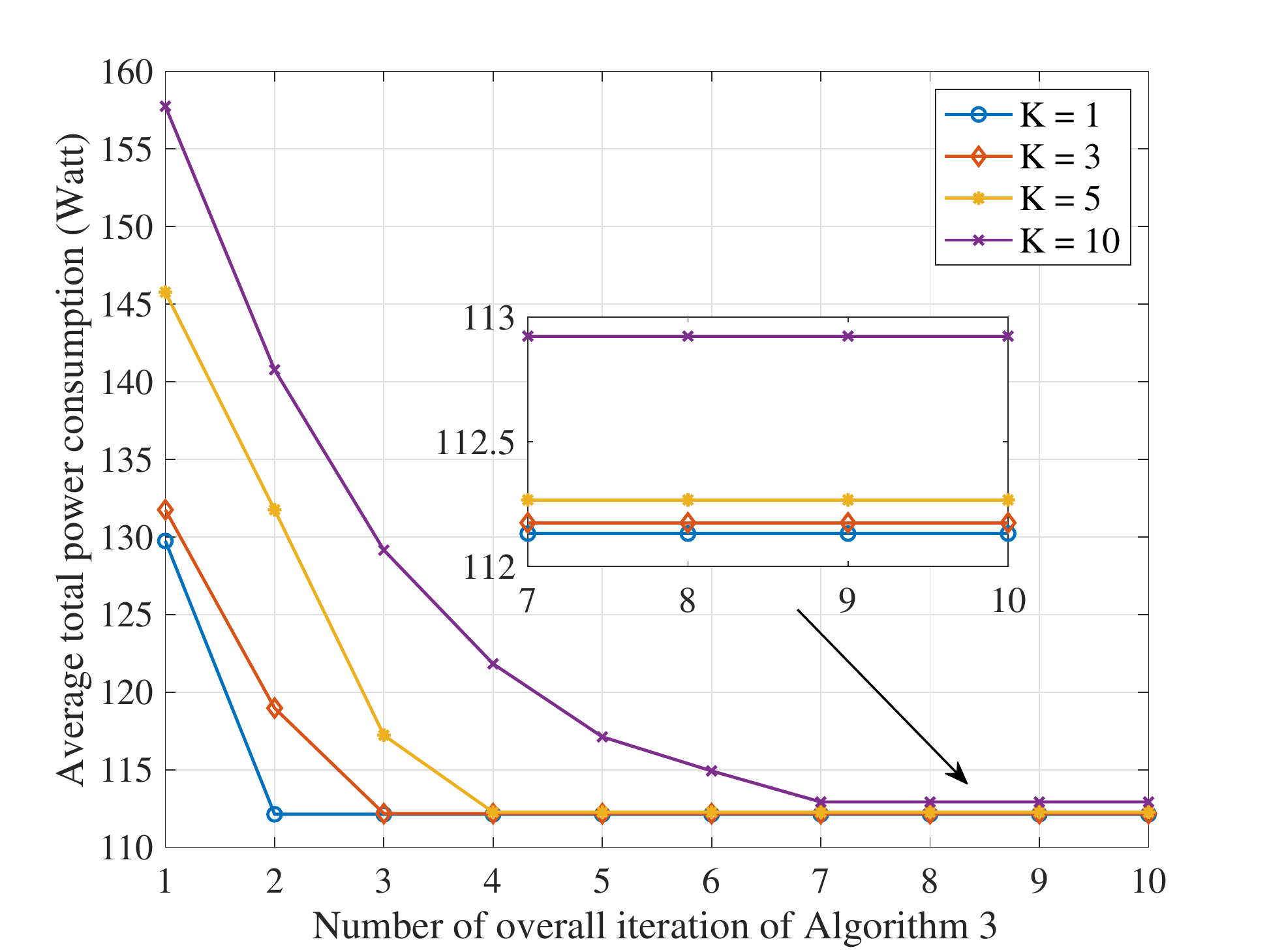}
      \vspace*{-9mm}
      \caption{Convergence of the PS for different number of GUs.}
      \label{converge_diff_k}
  \end{minipage}
  \vspace*{-9mm}
\end{figure}

{Fig. \ref{convergence} and Fig. \ref{converge_diff_k} illustrates the convergence behavior of the proposed alternating optimization algorithm in $\textbf{Algorithm \ref{alg_overall}}$ for minimizing the average total power consumption.
In order to compare the system performance of the PS with baseline schemes, we consider the PS with two different time durations, $T = 50$ s and $T = 100$ s in Fig. \ref{convergence}.}
In other words, there are $N = 500$ and $N = 1,000$ time slots in these settings, respectively.
Also, we set the minimum per GU required data rate as $R_{\min_k} = 3$ bits/s/Hz.
It can be observed that the system average total power consumption for the PS with different $T$ and $M_{\mathrm R}$ can rapidly converge to a suboptimal solution within only 5 iterations, which confirms the practicality of the proposed algorithm.
On the other hand, the NI scheme and CFA scheme enjoy a similar convergence rate as the PS but with worse performance.
The average total power consumption of the PCSI scheme converges to the lowest value among all the considered schemes since the PCSI scheme is the performance upper bound as perfect CSI is available which avoids outages and inefficient flight detour.
Detailed discussions comparing the PS and baseline schemes in terms of system performance and their corresponding trajectories will be presented in next sections.
{Moreover, as shown in Fig. \ref{converge_diff_k}, it can be observed that the PS converges quickly within 10 iterations for different number of GUs.
Also, more number of iterations are needed for the PS as the increase in the number of users enlarges the solution space substantially.
Therefore, based on the aforementioned theoretical discussion and simulation results, the proposed problem is still tractable while increasing the number of users.}
In the sequel, the maximum number of iterations of $\textbf{Algorithm \ref{alg_overall}}$ of the PS is set as 10 for illustration.

\vspace{-2mm}
\subsection{3D Trajectory of the UAV}
\vspace{-1mm}

Fig. \ref{vertical_view_trajectory} shows the bird's eye view of the UAV's trajectory obtained by the PS and baseline schemes with different setups.
In this figure, we set $R_{\min_k} = 3$ bits/s/Hz.
For the PS with a sufficiently long service time duration, i.e., $T = 100$ s, the UAV tends to maintain at a constant horizontal flight speed, i.e., $11$ m/s, as indicated by the spaces between two consecutive simulation points, to reduce the total power consumption at the expense of longer flight duration.
In contrast, for the case of short service time duration, i.e., $T = 50$ s, the UAV quickly flies over the service area with a relatively high velocity as there is insufficient time for adopting a slow speed or a long detour.
In such cases, since reducing the communication distances between the UAV and the desired GUs are not always possible, the PS would also increase the transmit power to satisfy the minimum individual data rate leads to high system power consumption.

\begin{figure}[t]\vspace*{-2mm}
	\centering
	\includegraphics[width=3.3 in]{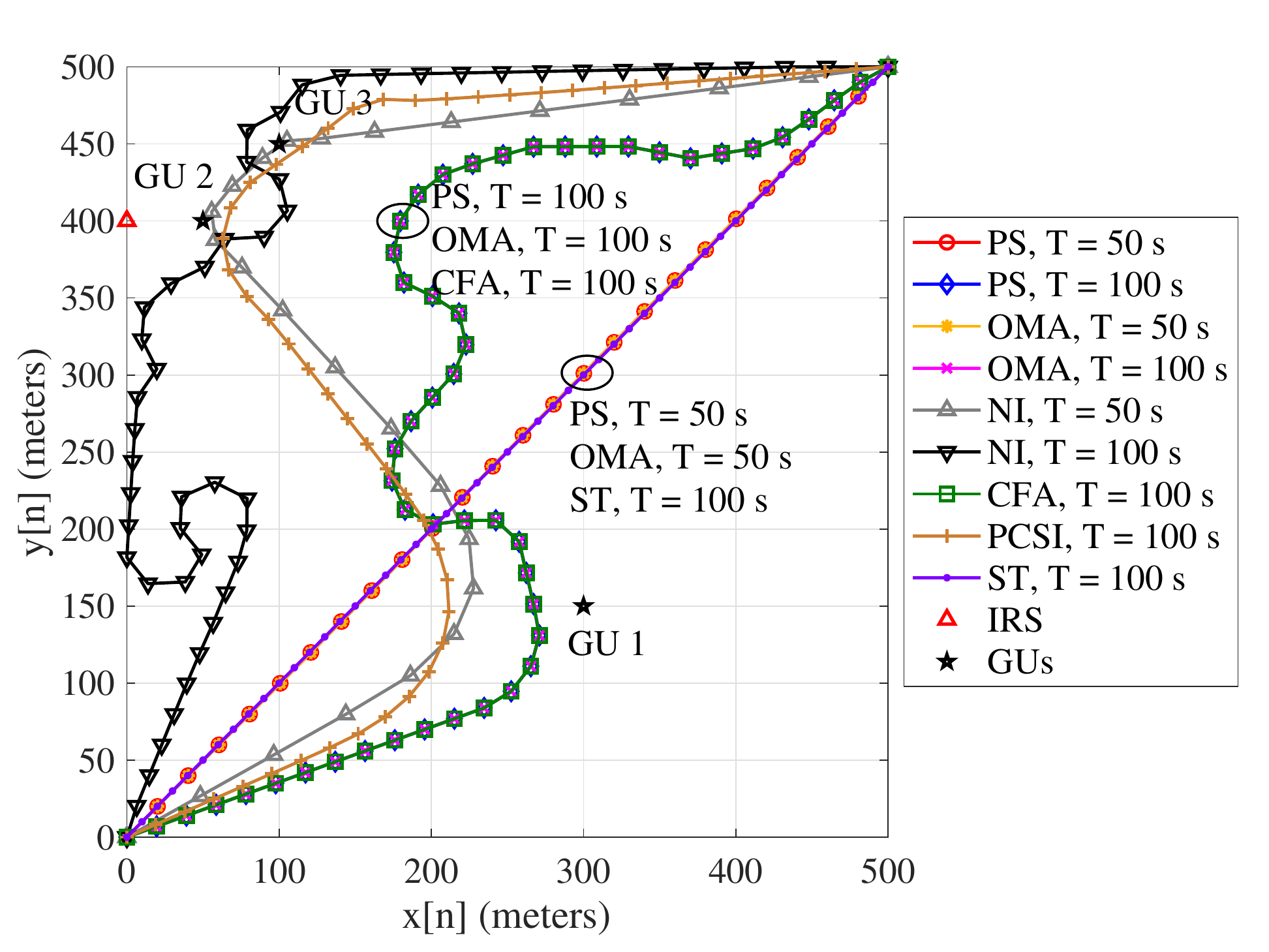}
	\vspace*{-6mm}
	\caption{The bird's eye view of the locations of the GUs and the IRS as well as the trajectory of the UAV for the PS and baseline schemes with different setups.}
	\label{vertical_view_trajectory}
	\vspace*{-9mm}
\end{figure}

For comparison, we also plot the UAV's trajectories for baseline schemes in Fig. \ref{vertical_view_trajectory}.
For the OMA scheme, the UAV's horizontal trajectory is similar to the PS for both short and long service time duration, i.e., $T = 50$ s and $T = 100$ s, respectively, as both schemes can efficiently exploit the extra degrees of freedom (DoF) offered by the IRS to optimize the UAV's trajectory.
For the NI scheme with a sufficiently long service time duration, i.e., $T = 100$ s, the UAV first flies towards GU 2 and GU 3 since these two users are close to each other creating a bottleneck in the system performance due to their minimum individual data rate constraints.
When the UAV is on the way to GU 2 and GU 3, the UAV would first deviate from the direct path to the centroid formed by GU 2 and GU 3 and fly towards GU 1 for communication such that it can effectively serve GU 1 to satisfy its data rate requirement.
Besides, the UAV would spend a sufficient number of time slots on GU 1 circling at the beginning of flight with a large transmit power to satisfy minimum data rate requirement of GU $1$ before approaching GU 2 and GU 3.
Thus, the UAV does not require to fly close enough to GU 1 to establish good channel conditions.
Moreover, for a shorter service time duration (i.e., $T = 50$ s) of the NI scheme, due to the insufficient number of time slots, the UAV has to fly with an exceedingly high flight velocity, on average $27$ m/s, and approach each GU to establish a strong gain channel for fulfilling the minimum date rate requirement for each GU.
In fact, this trajectory consumes a significantly high flight power due to the longer trajectory and higher flight velocity of the UAV.
Also, the UAV's 2D trajectory for the CFA scheme is the same as that for the PS with the same setups, e.g. $T = 100$ s, since the only differences between these schemes is whether to optimize the vertical dimension of the UAV or not.
As for the PCSI scheme, with a sufficient service time duration, i.e., $T=100$ s, the UAV approaches closely to each GU to satisfy the individual minimum data rate requirement with the most power-efficient flight velocity, i.e., $11$ m/s with the current setting, to effectively reduce the total system power consumption.
On the other hand, the ST scheme shares a similar route as the PS for the case of $T = 50$ s, which is the shortest path between the starting point and the destination.
However, the PS consumes much less system power consumption than that of the ST scheme, as will be shown in Fig. \ref{R_min}.

\begin{figure}[t]\vspace*{-2mm}
  \centering
  \begin{minipage}[t]{0.49\textwidth}
      \centering
      \hspace*{-5.3mm} \includegraphics[width=3.3 in]{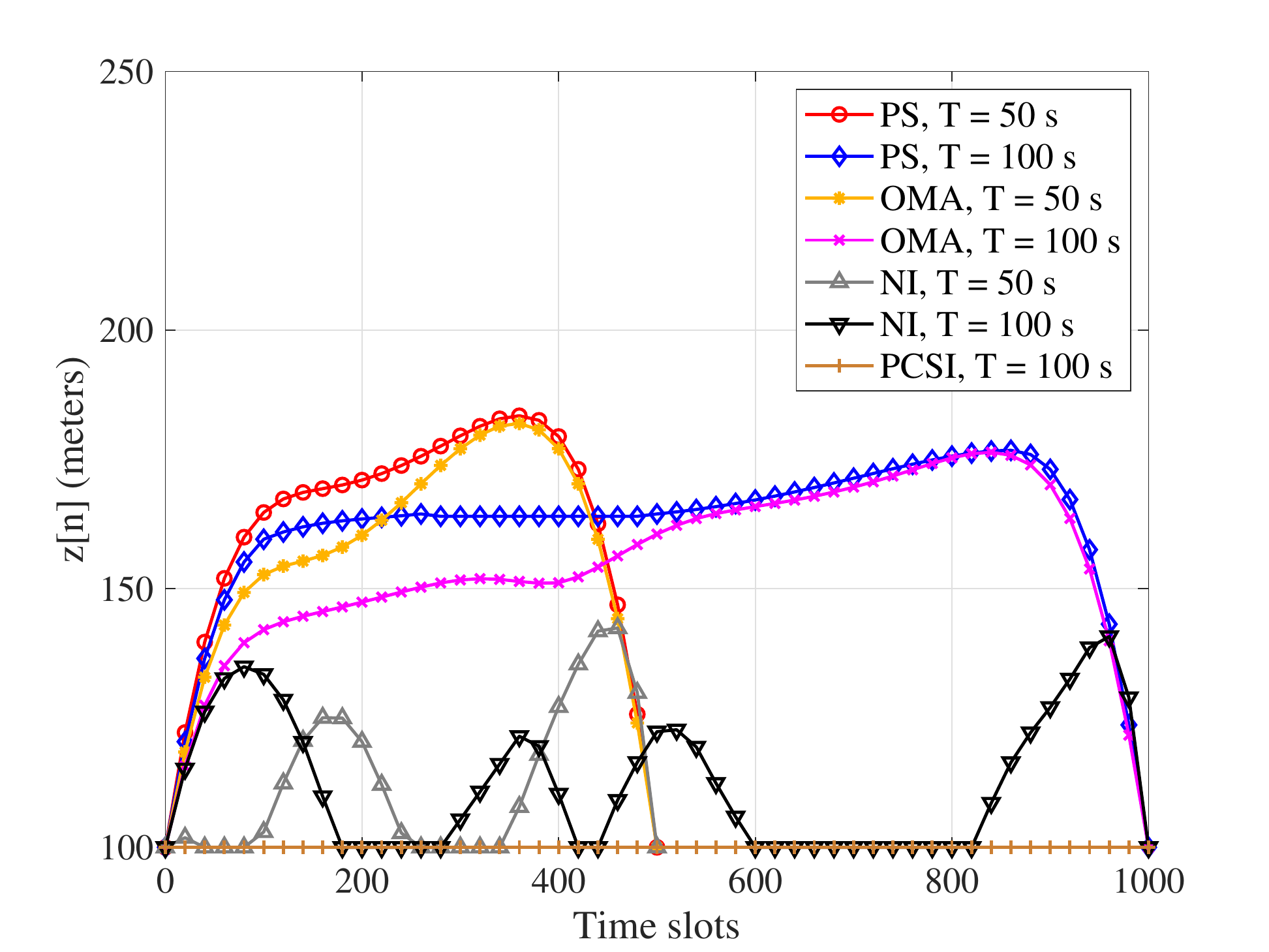}
      \vspace*{-9mm}
      \caption{The flight altitude of the UAV for the PS and the baseline schemes with different setups.}
      \label{height}
  \end{minipage}\,\,
  \begin{minipage}[t]{0.49\textwidth}
      \centering
      \hspace*{-6mm} \includegraphics[width=3.3 in]{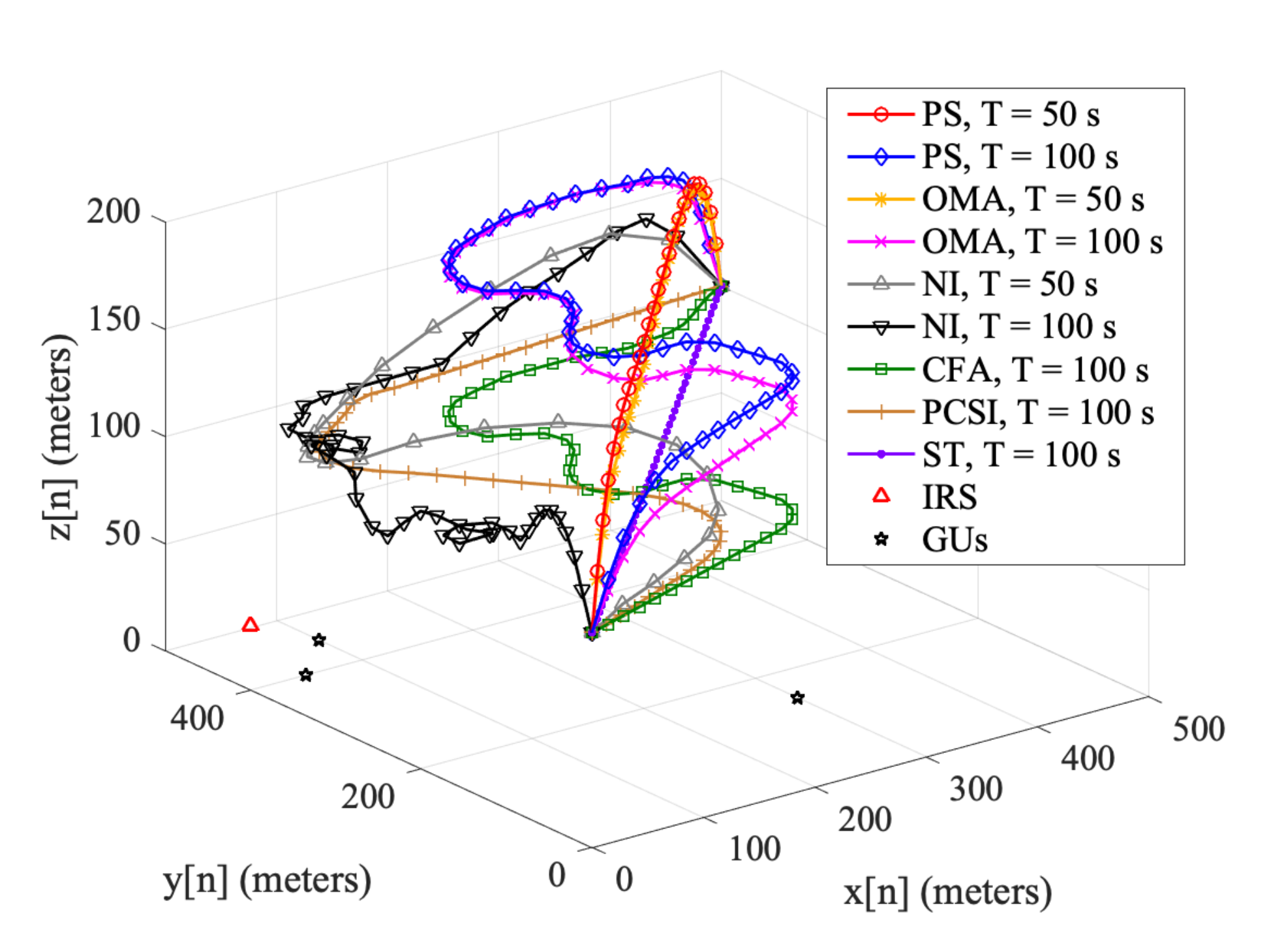}
      \vspace*{-9mm}
      \caption{The 3D view of the locations of the GUs and the IRS as well as the trajectory of the UAV for the PS and baseline schemes with different setups.}
      \label{3D_view}
  \end{minipage}
  \vspace*{-9mm}
\end{figure}

Fig. \ref{height} demonstrates the flight altitude of the UAV's trajectory for the PS and baseline schemes with different setups.
Since the UAV for the CFA scheme and the ST scheme have to flight at a constant altitude of $100$ m, we do not analyze the performance of these two baseline schemes in this figure.
For the PS with the two considered setups, i.e., $T=50$ s and $T=100$ s, the UAV prefers a high altitude with an optimized velocity in the journey to fully utilize the higher outage-guaranteed effective channel gain, c.f. Fig. \ref{comparisons_z}, since a higher outage-guaranteed effective channel gain can be obtained by adopting a moderately higher flight altitude when the UAV is far away from the GU in terms of horizontal distance.
Also, the flight altitude of the UAV adopting NOMA protocol is generally higher than the one adopting OMA.
Indeed, a higher altitude generally provides more freedom to the UAV to promote channel gain disparities of the selected two users for improving the performance of NOMA.
On the other hand, as OMA does not have the DoF in serving multiple users at each time instant, the UAV flying with a low to moderate altitude is good enough for it to strike a balance between data rate and outage probability.
{However, for the NI scheme, without the assistance of the IRS, the effective channel gain of the desired GUs is much lower} {than that of the PS.
Indeed, maintaining high-quality channels by reducing the path loss between the UAV and the selected GUs remains the key to satisfy the minimum data rate constraint in the NI scheme.}
Nevertheless, the UAV of the NI scheme is still willing to adopt a higher altitude occasionally to strike a balance among the total power consumption, outage-guaranteed effective channel gain, and path loss.
{In fact, as shown in Fig. \ref{height}, the parabolic patterns of the UAV's trajectory of the NI scheme appears in those time slots when the UAV is far away from any GUs, since in these locations, the outage-guaranteed effective channel gains are larger when the UAV operates at a higher altitude.}
In contrast, for the PCSI scheme, the UAV's flight altitude remains at the lowest possible altitude of $100$ m.
In fact, there is no channel outage event as the CSI is perfect known.
Thus, the UAV does not have any incentive to maintain a higher altitude as it would only consume more system energy but leading to a lower data rate.
To offer a better visualization of the trajectory of the PS and the baseline scheme, we also plot its 3D trajectory in Fig. \ref{3D_view}.
It can be seen from the optimized 3D trajectory that except the PCSI scheme, to effectively combat channel outages, the UAV should adopt a relatively high flight altitude to reduce the channel uncertainty caused by the altitude-dependent Rician fading, which not only reducing the communication power but also reducing the flight power of the UAV while achieving the same channel conditions.

\begin{figure}[t]\vspace*{-2mm}
  \centering
  \begin{minipage}[t]{0.49\textwidth}
      \centering
      \hspace*{-4.8mm} \includegraphics[width=3.3 in]{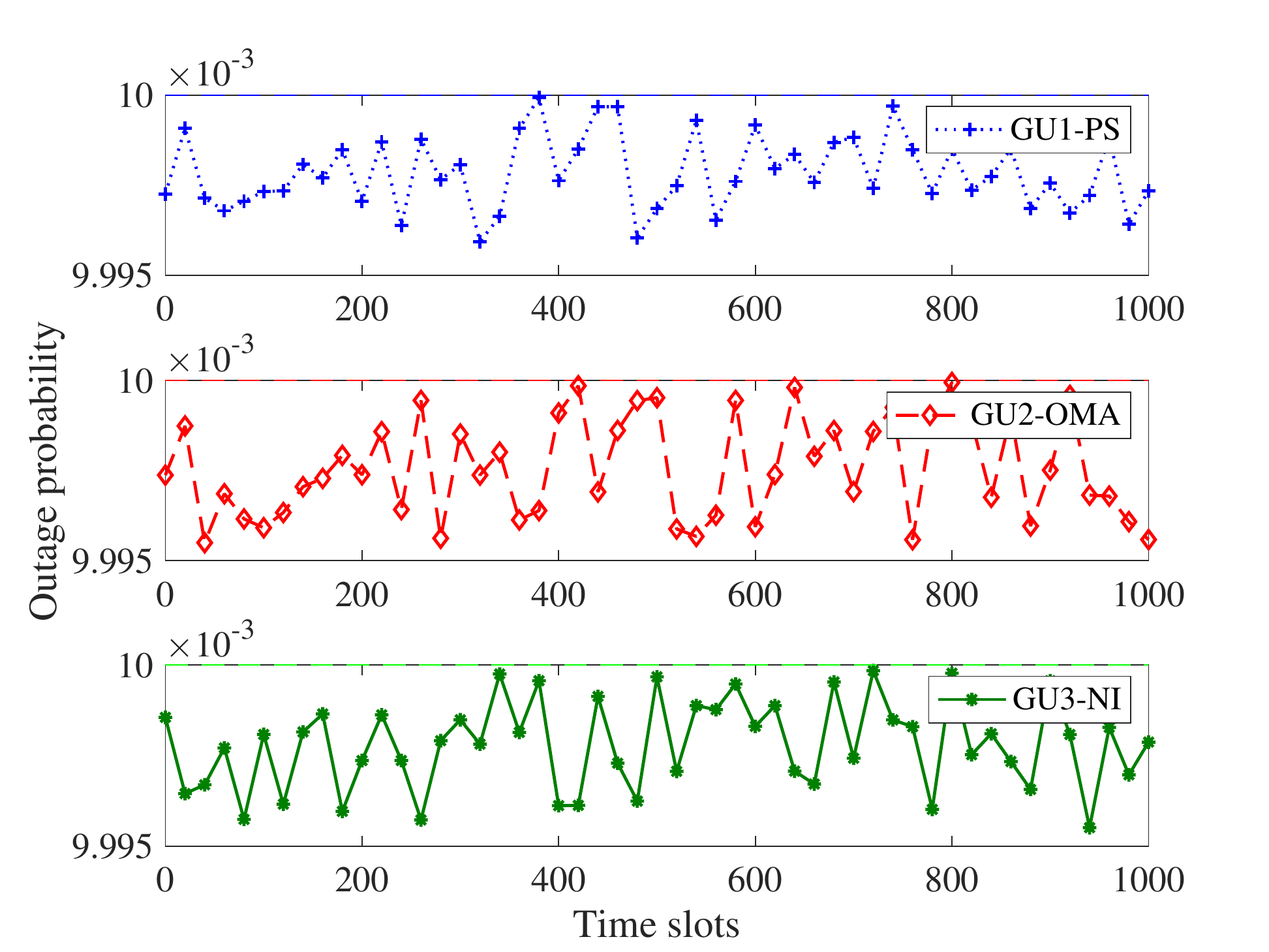} \vspace*{-9mm}
      \caption{The outage probability versus the time slots for GUs of the PS and two baseline schemes for $T=100$ s.}
      \label{outage}
  \end{minipage}\,\,
    \begin{minipage}[t]{0.49\textwidth}
      \centering
      \hspace*{-4.2mm} \includegraphics[width=3.3 in]{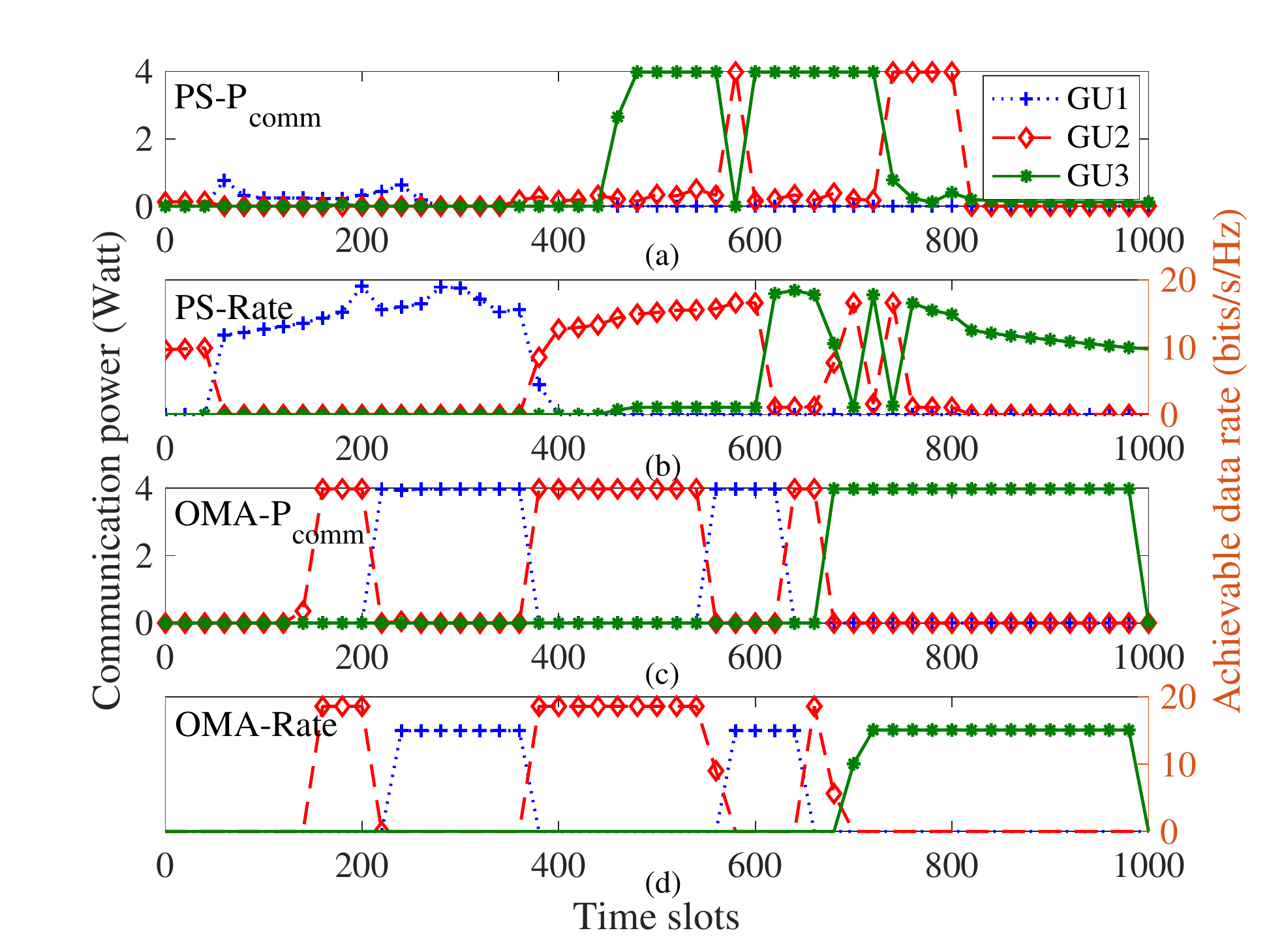}
      \vspace*{-9mm}
      \caption{The communication power consumption and the achievable data rate versus the time slots for each GUs of the PS and the baseline scheme for $T = 100$ s.}
      \label{comm_power_rate}
  \end{minipage}
  \vspace*{-9mm}
\end{figure}

\vspace{-1mm}
\subsection{Outage Probability}
\vspace{-1mm}

{
Fig. \ref{outage} demonstrates the outage probability versus the time slots for GUs of the PS and the baseline schemes for $T=100$ s.
We take the PS, OMA, and NI with $T=100$ s as examples }{ to calculate the outage probability as stated in constraint $\mathrm{C5}$.}
The outage probabilities in Fig. \ref{outage} were averaged over 1,000 random channel realizations by comparing the actual effective channel in \eqref{optimal_effective_channel} with the outage-guaranteed effective channel gain in \eqref{neural_network_model}.
{Thanks to the proposed DNN approach, the outage probability not only satisfies the required values, but also is close to its upper bound value, i.e., $\varepsilon = 0.001$, for any GU and time slots.
This illustrates the effectiveness of the DNN approach to approximate the outage-guaranteed effective channel gain and to be used for resource allocation design.}

\vspace{-1mm}
\subsection{Communication Power Consumption}
\vspace{-1mm}

{Fig. \ref{comm_power_rate} illustrates the communication power consumption and the achievable data rate versus the time slots for each GUs of the PS and baseline schemes for $T = 100$ s.
As shown in the sub-figures for the communication power (left hand side y-axis) and the achievable data rate (right hand side y-axis) of the PS, the UAV serves GU 2 and GU 3 from time slots $n = 420$ to $n = 830$, simultaneously, via the NOMA protocol.
Note that for those time slots adopting NOMA, the UAV allocates a significantly large portion of the communication power to the weak channel user to satisfy the corresponding minimum individual data rate requirement while a small power is allocated to the user with good channel condition.
This power allocation mechanism aligns with the one in the literature \cite{Zhiqiang_DC,sun2018robust,wei2018multi}.
In contrast, the total communication power consumption for the OMA scheme is much higher than the PS due to the less flexibility of the resource allocation.}

\vspace{-1mm}
\subsection{Average Total Power Consumption}
\vspace{-1mm}

{Fig. \ref{number_of_user} shows the average power consumption versus the number of GUs for the PS and the baseline schemes with different setups. 
In this simulation, we vary the number }{ of GUs, $K$, from 1 to 10 to illustrate the impact of the number of GUs on the system performance.
We set the locations of these GUs in x-dimension and y-dimension as $x_k = [300; 50; 100; 200; 150; 400; 100; 250; 300; 100]$ and $y_k = [150; 400; 450;100; 350; 400; 250; 250;$ $400; 50]$, respectively.
Besides, we assume that the minimum individual data rate is $R_{\min,k} = 3$  bits/s/Hz in this section.
It can be observed that the average power consumption of the PS increases with the number of the GUs as the system becomes less flexible in allocating resources when there are more numbers of GUs imposing more stringent QoS constraints.
Besides, for the PS with different numbers of IRS elements and time durations, the total power consumption of the system has only a marginal increase when the number of GUs is $K \geq 2$.
This can be attributed to the fact that the proposed optimization framework can achieve a better utilization of the system resources for serving a large number of GUs via jointly optimizing the UAV's 3D trajectory, IRS passive beamforming, and resource allocation.
Besides, when $K = 1$, the power consumptions of the PS and the baseline schemes OMA and NI are roughly the same since GU 1 is located far away from the IRS.
This result illustrates that the performance gain brought by the IRS is sensitive to its distances to the desired GUs.
In contrast, although the average power consumption for the NI scheme and the CFA scheme have a similar trend as the PS w.r.t. the number of GUs, the former two schemes consume a higher power than that for the PS under the same setting.
Indeed, the power consumption differences between these two schemes illustrate the performance gain brought by the IRS and the benefits of optimizing the UAV's flight altitude.
Also, the PCSI scheme is a performance upper bound of the PS which illustrates the influence of the altitude-dependent Rician fading channel to the proposed problem.}

\begin{figure}[t]\vspace*{-2mm}
\centering
\begin{minipage}[t]{0.49\textwidth}
   \centering
   \hspace*{-4.8mm} \includegraphics[width=3.3 in]{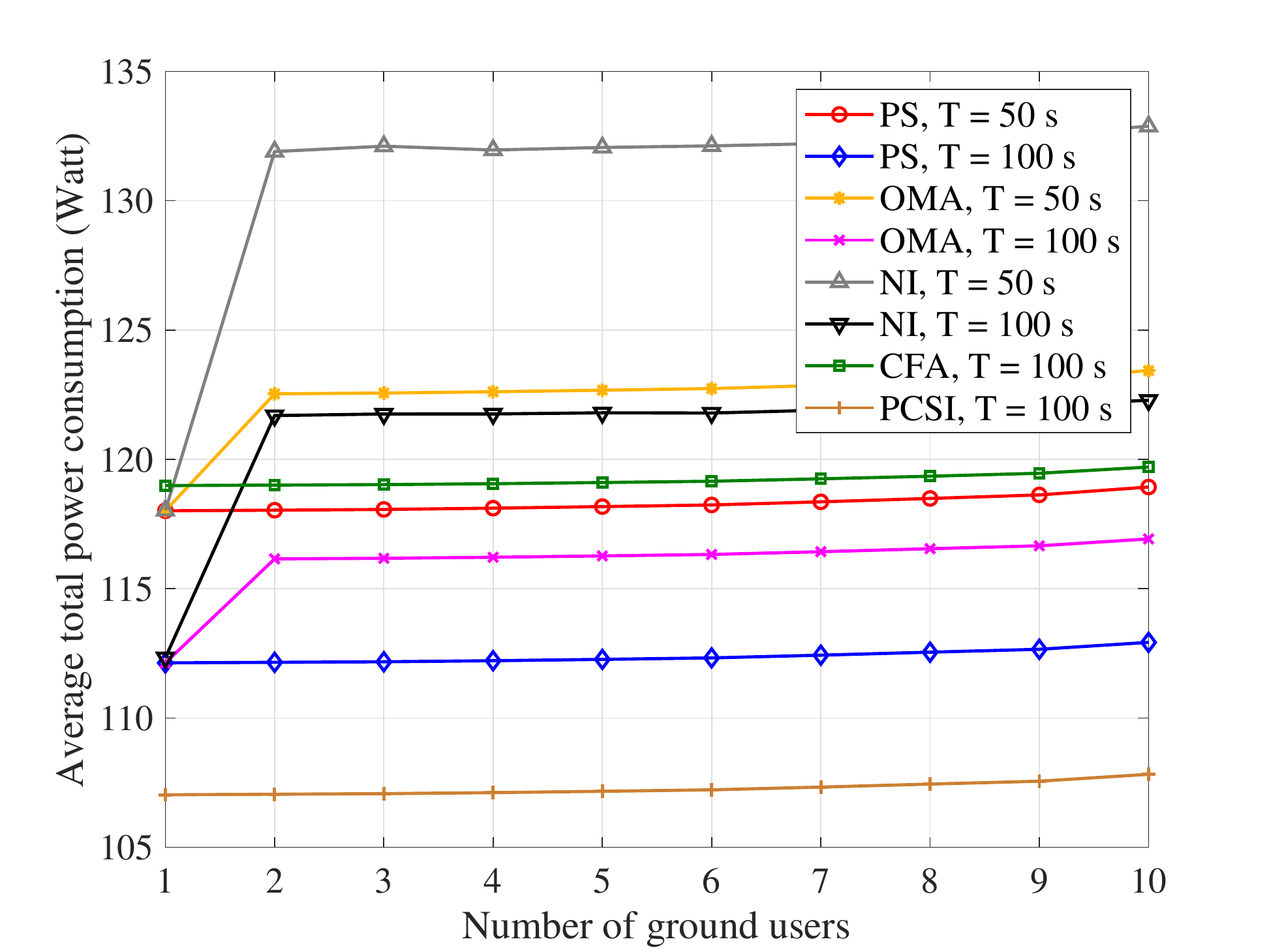} \vspace*{-9mm}
   \caption{The average total power consumption versus the number of GUs for the PS and the baseline schemes with different setups. }
   \label{number_of_user}
\end{minipage}\,\,
\begin{minipage}[t]{0.49\textwidth}
   \centering
   \hspace*{-4.2mm} \includegraphics[width=3.3 in]{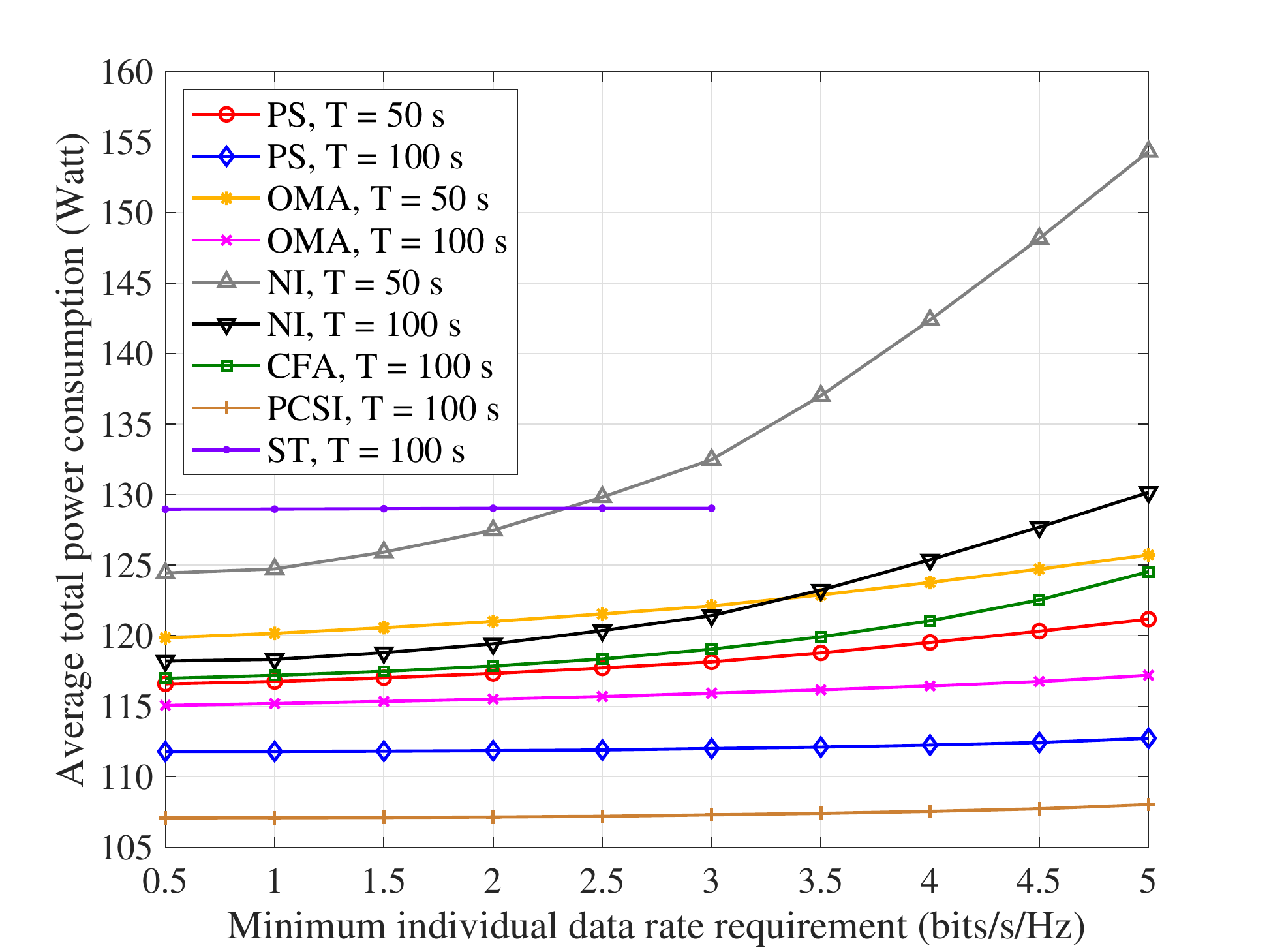} \vspace*{-9mm}
   \caption{The average power consumption versus minimum individual data rate requirement for the PS and baseline schemes with different setups.}
   \label{R_min}
\end{minipage}
\vspace*{-9mm}
\end{figure}

Fig. \ref{R_min} depicts the average total power consumption of the considered system versus the minimum individual data rate requirement for the PS and baseline schemes.
For the PS, the average power consumption slowly increases with the minimum individual data rate requirement compared with baseline schemes, since the IRS-assisted system can effectively optimize the system resources to minimize the average power consumption via optimizing the resource allocation and UAV's 3D trajectory.
In contrast, the average power consumption of the NI scheme and the CFA scheme scales with the minimum data rate requirement much faster than that of the PS.
The reason is that both the NI and CFA systems do not have sufficient DoF to optimize the system resources for the minimization of the total power consumption.
In particular, the former is due to lack of the contribution of the large number of elements equipped at the IRS and the latter is due to the fixed UAV's flight height.
As for the PCSI scheme, the average power consumption scales slowly with the minimum individual data rate requirement since the UAV's trajectory and resource allocation can be efficiently optimized due to the assistance of the IRS, which is similar to the PS.
As for the ST scheme, the average power consumption remains almost a constant between $R_{\min}=0.5$ bits/s/Hz and $R_{\min_k}=3$ bits/s/Hz.
Besides, since there are insufficient DoF and flexibility for optimizing the system resources, an exceedingly stringent minimum data rate requirement, i.e., $R_{\min_k}\geq 3.5$ bits/s/Hz, would lead to an infeasible result, which are not plotted in the figure.

\vspace{-2mm}
\section{Conclusions}
\vspace{-1mm}

In this paper, we minimized the average total power consumption in an IRS-assisted UAV-NOMA communication system via jointly optimizing the communication resource allocation, the 3D trajectory design of the UAV, and the phase shift control of the IRS.
The proposed formulation was a non-convex optimization problem taking into account the minimum outage probability and the minimum achievable data rate.
To handle the intractability of the outage constraint, we approximated the effective channel function via exploiting the DNN approach to facilitate an outage-guaranteed effective channel gain.
A suboptimal solution was achieved by the proposed iterative algorithm based on the alternating optimization method.
Numerical results illustrated that the proposed algorithm converges within a small number of iterations and revealed some interesting insights.
Particularly, (1) deploying an IRS to assist the UAV communication serves as a key to improve the system performance when the total service time is insufficient;
(2) employing the IRS-assisted UAV communication system offers enhanced flexibility in designing the UAV's trajectory;
{(3) optimizing the 3D trajectory of the UAV can strike a balance between the communication path loss and the altitude-dependent outage probability for improving the system power efficiency;
(4) NOMA communications offer a higher DoF for effective resource allocation design than that of the OMA scheme to minimize the average power consumption.}

\section{Appendix-Proof of Theorem \ref{optimize_phase_shift}}

In the formulated problem in \eqref{proposed_formulation_origion}, we can observe that the phase control policy of the IRS, $\bm \Phi$, only affects the distribution of $C_{k,k'}^{\mathrm{I}}[n]$, $C_{k,k'}^{\mathrm{II}}[n]$, $C_{k'\!,k}^{\mathrm{NSIC}}[n]$, and $C_{k,k}^{\mathrm{II}}[n]$ in constraints $\mathrm{C6}-\mathrm{C8}$, respectively.
Thus, for any given optimization variables $\mathcal{S}$, $\mathcal{P}$, $\mathcal{R}$, $\mathcal{T}$, and $\mathcal{V}$, a suboptimal $\bm \Phi$ can be obtained by maximizing the feasible probability of GU $k$ exploiting SIC decoding with stage II at $\mathrm{C6}$ in \eqref{proposed_formulation_origion}, e.g.\vspace{-1mm}
\begin{equation} \label{max_prob}
\underset{\bm \Phi}{\mathrm{maximize}} \,\, \mathrm{Pr} \left( s_{k,k'}[n] r_k[n] \leq s_{k,k'}[n] \log_2 \left( 1 + \frac{p_k[n] |h_{k}[n]|^2}{\sigma_k^2} \right) \right). \hspace{-5mm}\\[-1mm]
\end{equation}
Note that the effective channel follows $h_k[n]\sim\mathcal{CN} \left(\mu_h(\bm \Phi[n]), \sigma_h^2 \right)$, where $\mu_h(\bm \Phi[n])$ and $\sigma_h^2$ denote the mean and variance of the effective channel, respectively, such that $|h_k[n]|^2$ is noncentral chi-squared distributed.
Besides, the variance of the effective channel, i.e., $\sigma_h^2$, is independent of $\bm \Phi[n]$ as the introduced phase rotation does not change the distribution of the scattering component in \eqref{channel}.
Based on the CDF of $|h_k[n]|^2$ in \eqref{c5outage}, problem in \eqref{max_prob} can be rewritten as the following equivalent form\vspace{-1mm}
\begin{equation}\label{mininize_cdf}
\underset{\bm \Phi}{\mathrm{minimize}} \,\, F_{n,k} \left(\frac{\sigma_k^2 (2^{r_{k}[n]} -1)}{p_k[n]}, \lambda \right) = 1 - Q_{\varsigma} \bigg(\sqrt{\lambda}, \sqrt{\frac{\sigma_k^2 (2^{r_{k}[n]} -1)}{p_k[n]}} \bigg),\\[-1mm]
\end{equation}
where $\lambda=\frac{|\mu_h(\bm \Phi[n])|^2}{\sigma_h^2}$ and $Q_{\varsigma}(a,b)$ represent the noncentral parameter and the Marcum Q-function of the noncentral chi-square distribution with $2\varsigma$ DoF, respectively.
Moreover, it can be verified that the derivative of $F_{n,k}(\cdot,\lambda)$ w.r.t. $\lambda$ is less than 0 and problem \eqref{mininize_cdf} is equivalent to maximize the noncentral parameter $\lambda$ of $|h_k[n]|^2$, which is directly proportional to $|\mu_h(\bm \Phi[n])|^2$.
Therefore, the optimal phase control can be obtained by solving\vspace{-1mm}
\begin{equation}
\underset{\bm \Phi}{\mathrm{maximize}} \,\, |\mu_h \left(\bm \Phi [n] \right)|^2,\\[-1mm]
\end{equation}
where $\mu_h \left(\bm \Phi [n] \right)$ for a given $\bm \Phi [n]$ can be expressed as\vspace{-0.5mm}
\begin{align}
& \sqrt{\frac{\beta_0 \kappa_k^{\mathrm{AG}}[n]}{(d_k^{\mathrm{AG}}[n])^{\alpha^{\mathrm{AG}}}(1 + \kappa_k^{\mathrm{AG}}[n])}} e^{-j \frac{2 \pi d_k^{\mathrm{AG}[n]}}{ \lambda_{\mathrm c}}} + \sqrt{\frac{\beta_0^2 \kappa^{\mathrm{RG}}}{(d^{\mathrm{AR}}[n])^{\alpha^\mathrm{AR}} (d_k^{\mathrm{RG}})^{\alpha^{\mathrm{RG}}} (1+\kappa^{\mathrm{RG}})}} e^{-j \frac{2 \pi (d_k^{\mathrm{RG}}+d^{\mathrm{AR}}[n]) }{\lambda_{\mathrm c}}} \notag \\[-1mm]
& \times \big[ 1, e^{-j \frac{2 \pi \Delta_{\mathrm Rx}}{\lambda_{\mathrm c}} \sin \theta_{k}^{\mathrm{RG}} \cos \xi_{k}^{\mathrm{RG}}}, \ldots, e^{-j \frac{2 \pi \Delta_{\mathrm Rx}}{\lambda_{\mathrm c}} (M_{\mathrm Rx}-1) \sin \theta_{k}^{\mathrm{RG}} \cos \xi_{k}^{\mathrm{RG}}} \big] \notag \\[-1mm]
& \otimes \big[ 1, e^{-j \frac{2 \pi \Delta_{\mathrm Ry}}{\lambda_{\mathrm c}} \sin \theta_{k}^{\mathrm{RG}} \sin \xi_{k}^{\mathrm{RG}}}, \ldots, e^{-j \frac{2 \pi \Delta_{\mathrm Ry}}{\lambda_{\mathrm c}} (M_{\mathrm Ry}-1) \sin \theta_{k}^{\mathrm{RG}} \sin \xi_{k}^{\mathrm{RG}}} \big] \bm \Phi[n] \notag \\[-1mm]
& \times \big[ 1, e^{-j \frac{2 \pi \Delta_{\mathrm Rx}}{\lambda_{\mathrm c}} \sin \theta^{\mathrm{RA}}[n] \cos \xi^{\mathrm{RA}}[n]}, \ldots, e^{-j \frac{2 \pi \Delta_{\mathrm Rx}}{\lambda_{\mathrm c}} (M_{\mathrm Rx}-1) \sin \theta^{\mathrm{RA}}[n] \cos \xi^{\mathrm{RA}}[n]} \big]^{\mathrm H} \notag \\[-1mm]
& \otimes \big[ 1, e^{-j \frac{2 \pi \Delta_{\mathrm Ry}}{\lambda_{\mathrm c}} \sin \theta^{\mathrm{RA}}[n] \sin \xi^{\mathrm{RA}}[n]}, \ldots, e^{-j \frac{2 \pi \Delta_{\mathrm Ry}}{\lambda_{\mathrm c}} (M_{\mathrm Ry}-1) \sin \theta^{\mathrm{RA}}[n] \sin \xi^{\mathrm{RA}}[n]} \big]^{\mathrm H}.\\[-11mm] \notag
\end{align}
Note that maximizing the norm of the mean for the effective channel gain is equivalent to align the LoS component of the reflect link with that of the direct link.
Therefore, the suboptimal phase control policy of the IRS is obtained as in \eqref{phase_shift}.



\end{document}